\documentclass[11pt]{article}
\usepackage[total={6in, 8in}]{geometry}
\usepackage{amsthm,amsmath,amsfonts,amssymb}
\usepackage{graphicx}
\usepackage{bm}
\usepackage{threeparttable}
\usepackage{booktabs,multicol,multirow}
\usepackage{parskip}
\usepackage{authblk}
\usepackage{bm}
\usepackage{booktabs,multicol,multirow}
\usepackage{natbib}
\usepackage{float}

\title{A Sparse Factor Model for Clustering High-Dimensional Longitudinal Data}

\author[1,2,*]{Zihang Lu}
\author[3]{Noirrit Kiran Chandra}

\affil[1]{Department of Public Health Sciences, Queen’s University, Kingston, ON, Canada.}
\affil[2]{Department of Mathematics and Statistics, Queen's University, Kingston, ON, Canada.}
\affil[3]{Department of Mathematical Sciences, The University of Texas at Dallas, Richardson, Texas, USA.}
\affil[*]{Corresponding author: zihang.lu@queensu.ca}

\date{ }

\begin{document}

\maketitle

\raggedbottom

\begin{abstract}
Recent advances in engineering technologies have enabled the collection of a large number of longitudinal features. This wealth of information presents unique opportunities for researchers to investigate the complex nature of diseases and uncover underlying disease mechanisms. However, analyzing such kind of data can be difficult due to its high dimensionality, heterogeneity and computational challenges. In this paper, we propose a Bayesian nonparametric mixture model for clustering high-dimensional mixed-type (e.g., continuous, discrete and categorical) longitudinal features. We employ a sparse factor model on the joint distribution of random effects and the key idea is to induce clustering at the latent factor level instead of the original data to escape the curse of dimensionality.  The number of clusters is estimated through a Dirichlet process prior. An efficient Gibbs sampler is developed to estimate the posterior distribution of the model parameters. Analysis of real and simulated data is presented and discussed. Our study demonstrates that the proposed model serves as a useful analytical tool for clustering high-dimensional longitudinal data.
\end{abstract}

\section{Introduction}
In medical studies, longitudinal data are typically collected from the same individual or a group of individuals at multiple time points, allowing for the examination of changes and patterns over time.  Recent advances in engineering technologies have enabled the collection of large amounts of high-dimensional longitudinal data. This type of data refers to data that consists of measurements or observations taken over time on a large number of features or outcomes.  Examples include longitudinal and high dimensional multi-omic data \citep{Sailani2020},   imaging data \citep{Zipunnikov2014} and physical and cognitive status data \citep{Bruce1994}.   These features can be of different types, including continuous, discrete, categorical, or mixed-type, and can come from multiple sources such as high throughput sequencing, electronic health records and administrative databases. This wealth of information presents unique opportunities for researchers to investigate the complex nature of diseases and uncover underlying disease phenotypes. However, analyzing such data can be difficult due to their high dimensionality, heterogeneity and computational challenges. Statistical and machine learning methods for handling high dimensional longitudinal data have been emerging, for example, generalized estimation equations \citep{Wang2012},  linear mixed effect model \citep{An2013,Li2018},  test of significance \citep{Fang2020}, variable selection \citep{Fu2021}, supervised learning \citep{Adhikari2019, Capitaine2021}. \par

Although few methods are developed for clustering high-dimensional longitudinal data, there is a considerable amount of literature on clustering longitudinal data with low-dimensional features (also known as multi-dimensional longitudinal data) \citep{Lu2023}.  Examples include group-based multi-trajectory analysis \citep{Nagin2018},  K-means clustering \citep{Genolini2013, Bruckers2016},   Hierarchical clustering \citep{Zhou2023},  non-parametric clustering \citep{Lv2020},   model-based clustering using mixed effect models \citep{Marshall2006, Villarroel2009, Proust-Lima2015}   functional clustering\citep{Jacques2014, Bouveyron2011, Lim2020a},   Bayesian mixture models \citep{Neelon2011,  Lu2021, Lu2022, Tan2022a, Lu2023a}, and hidden Markov models \citep{Xia2019}, etc. These methods have been investigated only under a small number of features and their performance and scalability to the high dimensional setting have not been investigated.  In particular, \citet{Yang2022} proposed a model-based clustering method for high-dimensional longitudinal data via regularisation, and \citet{Sun2017} proposed a Dirichlet process mixture model for clustering longitudinal gene expression data. However, the former approach focuses on fixed and random effects selection on covariates while considering one single longitudinal outcome, while the latter approach requires the longitudinal gene data measured at a set of fixed time points that are identical between individuals.

In this paper, we aim to address this methodological gap by developing a novel Bayesian nonparametric mixture model for clustering high-dimensional, irregularly sampled and mixed-type longitudinal features. The proposed model assumes that the high-dimensional longitudinal data being collected are error-prone measurements of an unobserved low-dimensional set of latent variables (known as latent factors), and the key idea is to induce clustering at the latent factor level instead of the original data to escape the curse of dimensionality. The proposed model is appealing in several aspects: (a) the proposed model allows clustering of mixed-type (e.g., continuous, discrete and categorical) longitudinal features that are measured at irregular time points, (b) the number of measurements can differ between features and between individuals, (c) the number of clusters does not need to be specified a priori and is estimated from the model with a Dirichlet process prior, (d) an efficient Gibbs sampler is developed to estimate the posterior distributions of the model parameters. \par 

The rest of the paper is organized as follows: Section 2 describes the proposed Bayesian non-parametric mixture model for high dimensional longitudinal data.  Section 3 describes the Bayesian inference of the model parameters. Section 4 utilizes two case studies to demonstrate the utility of the proposed model.  Section 5 describes a simulation study to demonstrate the proposed model's performance compared to possible competing methods. Finally, Section 6 discusses our findings and future direction. 
 
\section{Methods}
In this section, we first briefly describe the latent factor model for clustering high dimensional cross-sectional data and then describe the proposed latent factor mixture for clustering high dimensional longitudinal data. 

\subsection{Latent factor mixture model}
The LAtent factor Mixture model for Bayesian clustering (lamb) originally proposed to cluster high-dimensional continuous data assuming that the intrinsic dimension of the data is small \citep{Chandra2023}.  Specifically, the lamb assumes that the high dimensional data being collected provide error-prone measurements of an unobserved low-dimensional set of latent variables (known as latent factors), and the clustering is performed at the latent factor level instead of the original data to achieve parsimony. \par

Let $\bm{y}_i = (y_{i1},...,y_{ip})^\top $ be a $p$-dimensional continuous response variables, for $i=1,...,N$.  A class of latent factor mixture models is defined as $\bm{y}_i \sim f(\bm{y}_i; \bm{\eta}_i, \phi) $ and 
$ \bm{\eta}_i \sim \sum_{h=1}^{\infty} \pi_h  g(\bm{\eta}_i; \bm{\theta}_h) $,  where $\bm{\eta}_i = (\eta_{i1},...,\eta_{id})^\top$ are $d$ dimensional latent variables. Also, $d < N$ and is assumed to be unknown.  The $f(\bm{y}_i; \bm{\eta}_i, \phi)$ is the density of the observed data conditional on the latent variables and parameter $\phi$. In addition, $g(\bm{\eta}_i; \bm{\theta}_h)$ is a $d$-dimensional kernel density with parameters $\bm{\theta}_h$, and $\bm{\eta}_i$ carries information of $\bm{y}_i$ at a lower dimensional space. The number of factors with non-negligible loadings is viewed as the active number of factors within each cluster.

As a canonical example, one can consider a Gaussian model with a mixture of Gaussians for the latent factor \citep{Chandra2023},   i.e.,
\begin{equation}
 \bm{y}_i \sim \text{MVN}_p(\bm{\Lambda}\bm{\eta}_i, \Sigma), \quad \bm{\eta}_i \sim \sum_{h=1}^{\infty} \pi_h \text{MVN}_d(\bm{\mu}_h, \bm{\Delta}_h), \quad  \pi_h \sim Q_0 
\end{equation} 
where $\text{MVN}_p$() denotes a $p$ dimensional multivariate normal distribution, $\bm{\Sigma} = \text{diag}(\sigma^2_1,...,\sigma^2_p)$ is a $p\times p$ diagonal matrix, $\bm{\Lambda}$ is a $p\times d$ matrix of factor loadings and $\bm{\theta}_h = (\bm{\mu}_h, \bm{\Delta}_h)$, where $\bm{\mu}_h$ and $\bm{\Delta}_h$ are mean and variance-covariance matrix of the multivariate normal distribution.   


\subsection{Latent factor mixture for clustering high dimensional longitudinal data}

In this subsection, we extend the lamb model to the longitudinal data setting. Let $\bm{y}_{ir}= (y_{ir1},...,y_{irn_{ir}})^\top$ denote the measurements for the $i$th individual and $r$th feature, where $i = 1,...,N$ and $r=1,...,R$.  Also, $n_{ir}$ is the number of measurements which can differ between $i$ and feature $r$. To account for different types of longitudinal features (i.e., continuous, discrete and categorical),  we assume the $ f_{r}(\bm{y}_{ir}|\bm{\gamma}_{r}, \bm{\beta}_{ir})$ is a distribution with a dispersion parameter $\vartheta_{r}$ and the fully specified mean function given by 
\begin{equation}
h^{-1}_{r}(E(\bm{y}_{ir}|\bm{\beta}_{ir},\bm{\gamma}_{r})) = \bm{\xi}_{ir} = \bm{x}_{ir}^\top\bm{\gamma}_{r} +  \bm{Z}_{ir}^\top\bm{\beta}_{ir}
\end{equation} 
where $h_{r}^{-1}$  is an element-wise canonical link function for the mean of the feature $r$. For example, an identify link function can be used for continuous data following Gaussian distribution, a log link function can be used for count data following Poisson distribution and a logit link can be used for binary data following Binomial distribution. The model also involves dispersion parameters $\vartheta_{r} = \sigma^2_{r}$, for Gaussian distribution. The corresponding dispersion parameters $\vartheta_r$ for Binomial and Poisson distributions are 1.  Also,  $\bm{x}_{ir} = (\bm{x}_{ir1}, ..., \bm{x}_{irp_r})^\top$ is a $p_r\times n_{ir}$ design matrix for fixed effect covariates, where $\bm{x}_{irm} = (x_{irm1}, ..., x_{irmn_{ir}})^\top$ for $m = 1,...,p_r$, where $p_r$ is the dimension of the fixed effect covariates. In practice, each longitudinal feature may share the same set of fixed effect covariates, and in such a case, $p_1 = p_2 = ... = p_R = p$. Furthermore, $\bm{\gamma}_{r}$ is the corresponding vector of coefficients, $\bm{Z}_{ir}$ is a $q_r\times n_{ir}$ design matrix for random effect covariates (e.g., time or age), and $\bm{\beta}_{ir} = (\beta_{ir1},..., \beta_{irq_r})^\top $ is the corresponding random effect coefficients, where $q_r$ is the dimension of the random effect for feature $r$. For identification reasons, we assume that the vectors  $\bm{x}_{ir}$ and $\bm{Z}_{ir}$  do not contain the same covariates. This specification is similar to the model proposed by \citet{Komarek2013}. 

To model the dependence within and between features, we assume a joint distribution between random effects estimated from different features. Specifically, let $\bm{\beta}_{i} = (\bm{\beta}_{i1}^\top,...,\bm{\beta}_{iR}^\top)^\top$ be the joint vector of random effect for all $R$ features of individual $i$. To induce clustering, one could propose to directly perform clustering on $\bm{\beta}_{i}$. However, the dimension of the joint random effect $\bm{\beta}_{i}$ grows rapidly with the number of features $R$ increasing and the dimension of the random effect $q_r$ for each feature $r$ increasing. This could cause an unstable model due to the curse of dimensionality. To mitigate this problem, we consider a sparse factor model \citep{Bhattacharya2011} on $\bm{\beta}_{i}$, in which the factor loading $\bm{\eta}_i$ is assumed to come from an infinite mixture model. Thus, the latent factor mixture model for the joint random effects is
\begin{equation}
\bm{\beta}_i \sim \text{MVN}_q(\bm{\Lambda}\bm{\eta}_i, \bm{\Sigma}_\beta), \quad \bm{\eta}_i \sim \sum_{h=1}^{\infty} \pi_h \text{MVN}_d(\bm{\mu}_h, \bm{\Delta}_h),  \quad \pi_h \sim Q_0
\end{equation} 
where $\bm{\Lambda}$ is a $q \times d$ factor loadings matrix with $d < q$,  $\bm{\eta}_i = (\eta_{i1},...,\eta_{id})^\top$ is a $d$-dimensional latent factor for individual $i$, and  $\bm{\Sigma}_\beta = \text{diag}(\omega_{\beta 1}^2,...,\omega_{\beta q}^2)$.  Equivalently, the model can be written as 
$ \bm{\beta}_i = \bm{\Lambda} \bm{\eta}_i + \bm{\epsilon}_i, \quad  \text{with}  \quad \bm{\epsilon}_i \sim \text{MVN}_q(\bm{0}, \bm{\Sigma}_\beta) $, where $\bm{\epsilon}_i = (\epsilon_{i1},...,\epsilon_{iq})^\top $,  for $i = 1,...,N$, is a residual vector that is independent with the other variables in the model and is normally distributed with mean zero and a diagonal covariance matrix  $\bm{\Sigma}_\beta$.  In practice, the latent factor dimension $d$ is unknown and needs to be set. In the current study, we choose the smallest $d$ (denoted by $\widehat{d}$) which explains at least 95\% of the random effect obtained from mixed effect models. The $\widehat{d}$ can be found using methods such as principal component analysis. A sketch of the proposed model framework is described in Figure 1. 

\begin{figure}[H]  
 \centering
   \includegraphics[width=14cm,height=5.5cm]{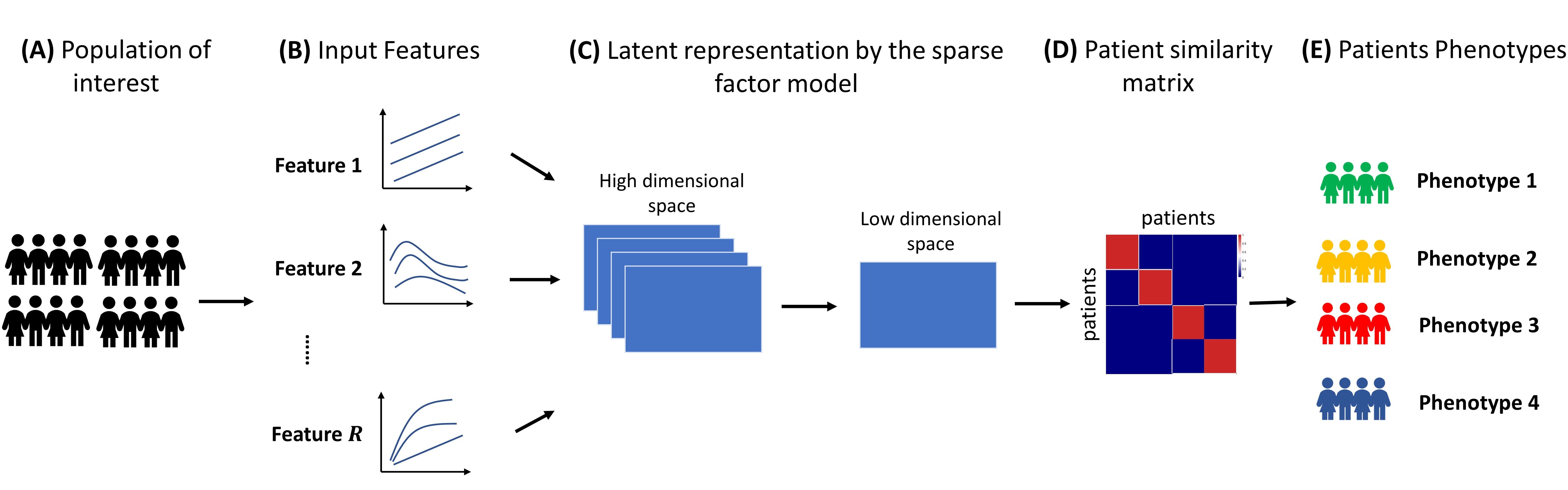}
\caption{Diagram of the proposed mixture of factor model for high-dimensional longitudinal data }
\end{figure}

\section{Bayesian Inference}
In this section, we describe the specification of prior distributions for the proposed model and sketch the computation steps for the posterior distribution. 

\subsection{Specification of the prior distributions}
Given the dimensionality, choosing conditionally conjugate priors that lead to efficient posterior computation via blocked Gibbs sampling is important. Here, we complete our proposed model by describing the specification of the prior distributions for model parameters. 

To induce shrinkage and reduce dimensionality, a Dirichlet-Laplace prior \citep{Bhattacharya2015} with parameter $a$ is used. Specifically, we let $\text{vec} (\bm{\Lambda}) \sim \text{DL}(a)$. Equivalently, let $\lambda_{kh}$ denote the element of $\bm{\Lambda}$ in the $k^{th}$ row and $h^{th}$ column, for $k=1,...,q$ and $h=1,..,d$, the hierarchal form the prior can be written as
\begin{equation}
\lambda_{kh} |\cdot \sim \text{N}(0, \varphi_{kh} \phi^2_{kh}\tau^2 ) , \quad \varphi_{kh} \sim \text{Exp}(1/2), \quad \phi_{kh} \sim \text{Dir}(a,...,a), \quad \tau \sim \text{Gamma}(qda, 1/2) 
\end{equation} 
where  $\text{Exp}(a)$ is an exponential distribution with mean $1/a$,  $\text{Dir}(a,...,a)$ is the $q$ dimensional Dirichlet distribution, and $\text{Gamma}(a,b)$ is a gamma distribution with mean $a/b$ and variance $a/b^2$. In addition, for the diagonal elements of $\bm{\Sigma}_\beta$, we set $\omega_{\beta k}^{-2} \sim \text{Gamma} (a_\omega, b_\omega)$, for $k=1,...,q$. For residual variance $\sigma^2_r$, we set $\sigma^{-2}_r \sim \text{Gamma} (a_{\sigma,r}, b_{\sigma,r})$, for $r = 1,...,R$.  For the cluster weight $\pi_h$, for $h=1,..., \infty$, we use a stick-breaking prior from a Dirichlet process, which has concentration parameter $\alpha $ impacting the induced prior on the number of clusters. A recent study shows that the consistency for the number of clusters in the Dirichlet process mixtures can be achieved if the concentration parameter is adapted in a fully Bayesian way \citep{Ascolani2023}. The hierarchical model defined in (3) for $\bm{\eta}_i$ can be represented as 
\begin{equation}
 \bm{\eta}_i |   \bm{\mu}_i, \bm{\Delta}_i \sim \text{MVN}_d(\bm{\mu}_i, \bm{\Delta}_i), \quad  \bm{\mu}_i, \bm{\Delta}_i |G \sim G, \quad  G \sim \text{DP}(\alpha, G_0), \quad \alpha \sim \text{Gamma} (a_\alpha, b_\alpha) 
\end{equation} 
where the base distribution $G_0 =\text{NIW}(\bm{\mu}_0, \bm{\Delta}_0, \kappa_0, \nu_0) $, where $\bm{\mu}_0$ and $\bm{\Delta}_0$ are mean vector and inverse scale matrix, respectively,  $\kappa_0$ and  $\nu_0$ are precision parameter and degree of freedom, respectively. 


\subsection{Posterior computation}
To obtain the posterior distribution of model parameters, we developed a Gibbs sampling algorithm with a split-merge update \citep{Jain2004}.   The split-merge update allows the algorithm to escape local modes and therefore resulting in an appropriate clustering of individuals. The detailed steps of sampling model parameters are described in Section A of the supplementary material. Convergence of the algorithm is diagnosed using visual inspection of trace plots and the Geweke statistics \citep{Geweke1991}.  The Geweke statistics determines whether a Markov chain convergence or not based on a test for equality of the means of the first and last part of a Markov chain (by default the first 10\% and the last 50\%). If the samples are drawn from the stationary distribution of the chain, the two means are equal and Geweke’s statistic has an asymptotically standard normal distribution.

\subsection{Inference of clusters}
In the mixture model, a well-known issue is the label-switching problem. This problem arises because the likelihood function of the model is invariant to the re-parametrization of the cluster labels. Due to this issue, the cluster labels and the cluster-specific parameters are not identifiable through posterior inference. However, the posterior probability of two individual $i$ and $j$ assigning to the same clusters $P(c_i = c_j|\bm{Y})$ is identifiable, where $c_i$ denotes the cluster label for individual $i$ and $\bm{Y}$ denotes data from all individuals. Using this approach, we first define a distance metric $\widetilde{d}(i,j) =  1 - P(c_i = c_j|\bm{Y})$, which reflects the probability of individual $i$ and $j$ not being assigned to the same cluster, and therefore it measures the dissimilarity between the two individuals. $\widetilde{d}(i,j) $ can be estimated by $\sum_{s=1}^SI(c_i = c_j)$, where $S$ is the number of MCMC iterations. A $N \times N$ posterior similarity matrix is constructed with $\widetilde{d}(i,j) $ as its $(i, j)$ element. Then the clustering is obtained by minimizing the posterior expectation of the Binders loss function \citep{Binder1978} .

\section{Application}
In this section, we apply the proposed model to two real-life examples to demonstrate its utility. The first dataset concerns analyzing high dimensional longitudinal cytokines data to identify distinct biological patterns for individuals living in California while the second dataset concerns identifying disease phenotypes of Primary Biliary Cholangitis (PBC) patients based on mix-typed biomarkers. 

\subsection{Example 1:  Longitudinal Cytokine Data}
\subsubsection{Data description}
The data come from a study of multi-omics seasonal patterns for a group of healthy individuals living in California \citep{Sailani2020}.  The original cohort includes longitudinal multi-omics data from profiling 105 individuals aged 25 to 75 years old. The data from this cohort includes quarterly sample collections of transcriptomes from peripheral blood mononuclear cells, proteome and metabolome from plasma and targeted cytokine and growth factor assays using serum. In total, there were 902 visits (average across different types of omics) from which samples were collected over up to 4 years.  These samples were then aggregated and mapped to a 1-year-long time frame \citep{Sailani2020}.  The sample collections were generally evenly distributed throughout the year. However, the measurement time points and frequencies differed between participants (Figure 2A). In addition, participants in this study were characterized for steady-state plasma glucose (SSPG), in which 31 participants were insulin sensitive (SSPG $<$ 150 mg/dL), and 35 were insulin resistant (SSPG $\ge$ 150 mg/dL). 

In the current analysis, we included 61 participants who have cytokine (immune protein) data. Cytokines are a broad category of small proteins that play a crucial role in cell signaling. They are primarily produced by immune cells and are involved in coordinating immune responses and regulating various biological processes in the body. Among the included participants, 27 (44.3\%) were insulin sensitive (IS) and 34 (55.7\%) were insulin resistant (IR) individuals (insulin resistance is often associated with Type 2 diabetes). The goal of the analysis is to identify individuals' subgroups based on 66 longitudinal cytokines. The full list of variable names is provided in Table S1. All cytokines were log-transformed prior to the cluster analysis. The individual trajectories of two selected cytokines (IL17F and MCSF) over time are shown in Figure 2B and Figure 2C, respectively. Subsequently, to further characterize the resulting clusters, we associated the clusters with individuals' insulin status. 

\begin{figure}[H]  
 \centering
   \includegraphics[width=13cm, height=8cm]{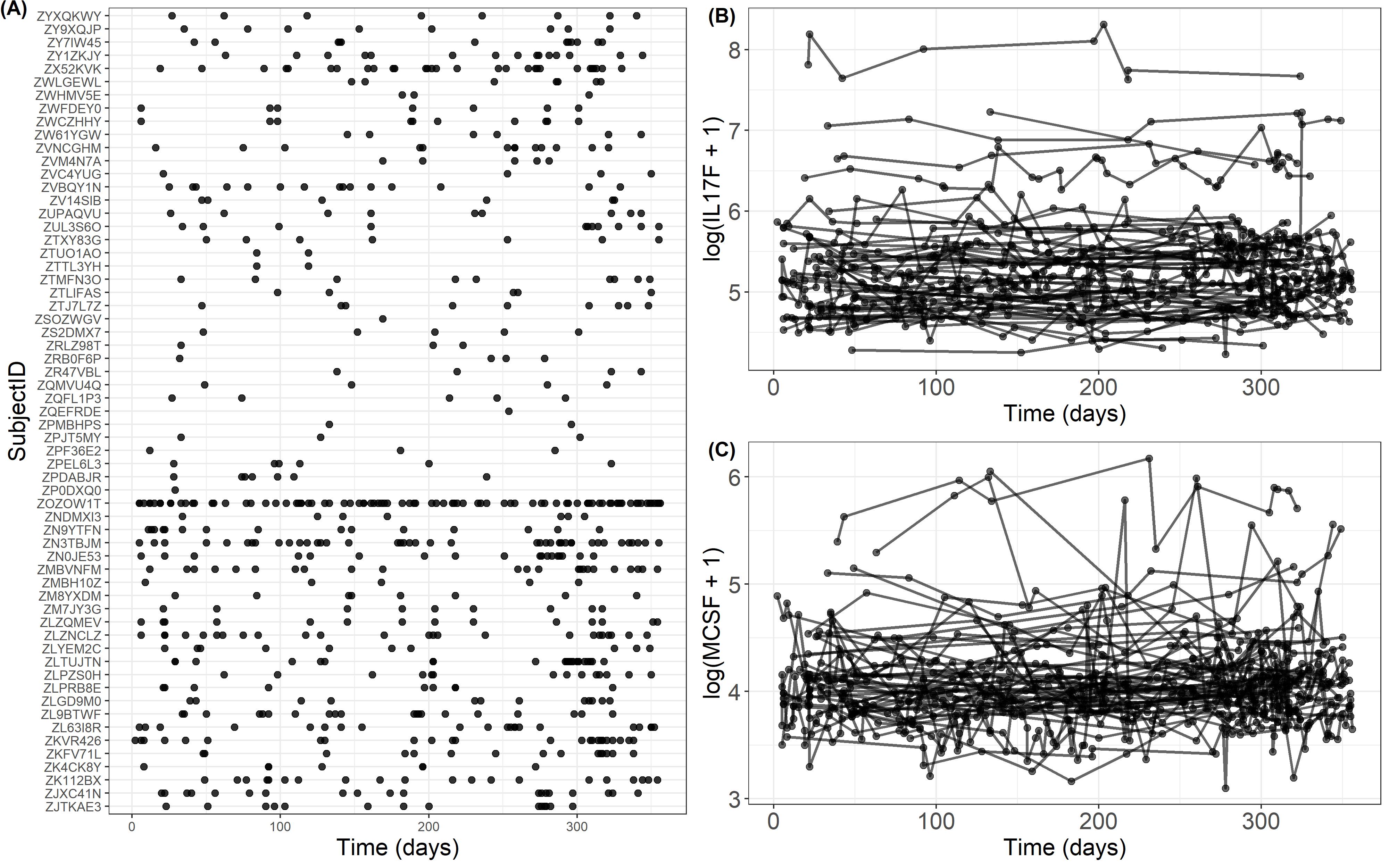}
\caption{Observed time grid for longitudinal immune protein data by individual ID and longitudinal trajectories of two selected immune protein markers for the longitudinal cytokines data (Example 1). (A) Observed time grid of each individual in the longitudinal immune protein data. (B) Longitudinal trajectory of IL17F in log scale. (C) Longitudinal trajectory of MCSF  in log scale. }
\end{figure}

\subsubsection{Model specification}
For this application, hyperparameters were set to reflect weak prior information for the model parameters. Specifically, following a previous study \citep{Bhattacharya2015}, we set $a = 0.5$ for the Dirichlet-Laplace prior distribution. Also, we set $a_{\sigma,r} = b_{\sigma,r} = 0.01$, $a_{\omega,r} = b_{\omega,r} = 0.01$ for $r = 1, ..., R$. For the concentration parameter, we considered four different scenarios, namely (a)  $a_{\alpha} = b_{\alpha} = 0.1$, reflecting a weakly informative prior (Figure 3A), (b) $a_{\alpha} = b_{\alpha} = 1$, reflecting a weakly to moderately informative prior (Figure 3B), (c) $a_{\alpha} = b_{\alpha} = 10$, reflecting a moderately to strongly informative prior (Figure 3C), and (d)  $a_{\alpha} = b_{\alpha} = 50$, reflecting a  strongly informative prior (Figure 3D). For the base distribution $G_0$, we set $\bm{\mu}_0 = \bm{0}$,  $\bm{\Delta}$ to be a diagonal matrix with elements being 80, $\nu_0 = 116 $ and $\kappa = 0.001 $. \par

To allow sufficient flexibility and to fully capture the dependence between markers and measurements within individuals, a linear form random effects of time (i.e., both random intercept and slope) was used, i.e., we set $\bm{Z}_{ir} =(1, {t}_{ir})^\top$, where ${t}_{ir} = (t_{i1},...,t_{in_i})^\top$.  This resulted in a dimension of 132 for random effect $\bm{\beta}_i$. In addition, the model is adjusted for a baseline covariate BMI, i.e., $\bm{x}_{ir} =(\text{BMI}_{ir})^\top$ for $r=1,...,R$. We ran the model with 20000 iterations, discarded the first 10000 samples and kept every 10th sample. This resulted in 1000 samples for each model parameter.

Next, we used leave-one-out cross-validation to investigate the cluster stability.  We created 61 subsets from the original data by holding out 1 individual \citep{Hennig2007}. We refit the proposed model based on this subset and compared the results (both the estimated number of clusters and the individual cluster membership) to those obtained from the original data. The agreement between the two clusterings was measured using the adjusted rand index (aRand) \citep{Hubert1985}.  A higher aRand suggests a better agreement between the two clusterings under evaluation. 

\subsubsection{Analysis results}

The clustering results based on the model under different values of concentration parameter $\alpha$ for the Dirichlet process prior are shown in Figure 3. The trace plots for $\alpha$ showed that all models were mixing well (Figure S1).  All models consistently identified two clusters (Figure 3E, F, G and H) and the clusterings were quite stable even for extreme choice of prior hyperparameters, for example, $\alpha \sim \text{Gamma}(0.1, 0.1)$ and $\alpha \sim \text{Gamma}(50, 50)$, suggesting that the resulting clustering uncovered the underlying heterogeneity that may reflect a biological difference among these individuals. The distribution of the clusters differed slightly under different prior of $\alpha$ (Figure 3I, J, K, L). For example, when $\alpha \sim \text{Gamma}(0.1, 0.1)$, 7 (11\%) individuals were assigned to Cluster 1 while 54 (89\%) individuals were assigned to Cluster 2. When $\alpha \sim \text{Gamma}(50, 50)$, 12 (20\%) individuals were assigned to Cluster 1 while 49 (80\%) individuals were assigned to Cluster 2. The resulting clusters showed distinct prevalence in individuals' insulin status, further suggesting biological differences between the clusters (Figure 3M, N,O, P). For example, under the model with $\alpha \sim \text{Gamma}(0.1, 0.1)$, all 7 individuals (100\%) from Cluster 1 were insulin sensitive whereas only 20( 37\%) individuals from Cluster 2 were insulin sensitive. Under the model with $\alpha \sim \text{Gamma}(50, 50)$, 9 (75\%) individuals from Cluster 1 were insulin sensitive whereas only 18 (36.7\%) individuals from Cluster 2 were insulin sensitive. The leave-one-out validation showed that the median number of clusters over 61 data subsets was around 2 for all models (Figure 4A) while the model with $\alpha \sim \text{Gamma}(50, 50)$ yielded slightly higher aRand, suggesting clustering were more stable compared to other models.

\begin{figure}[H]  
 \centering
   \includegraphics[width=13cm, height=13cm]{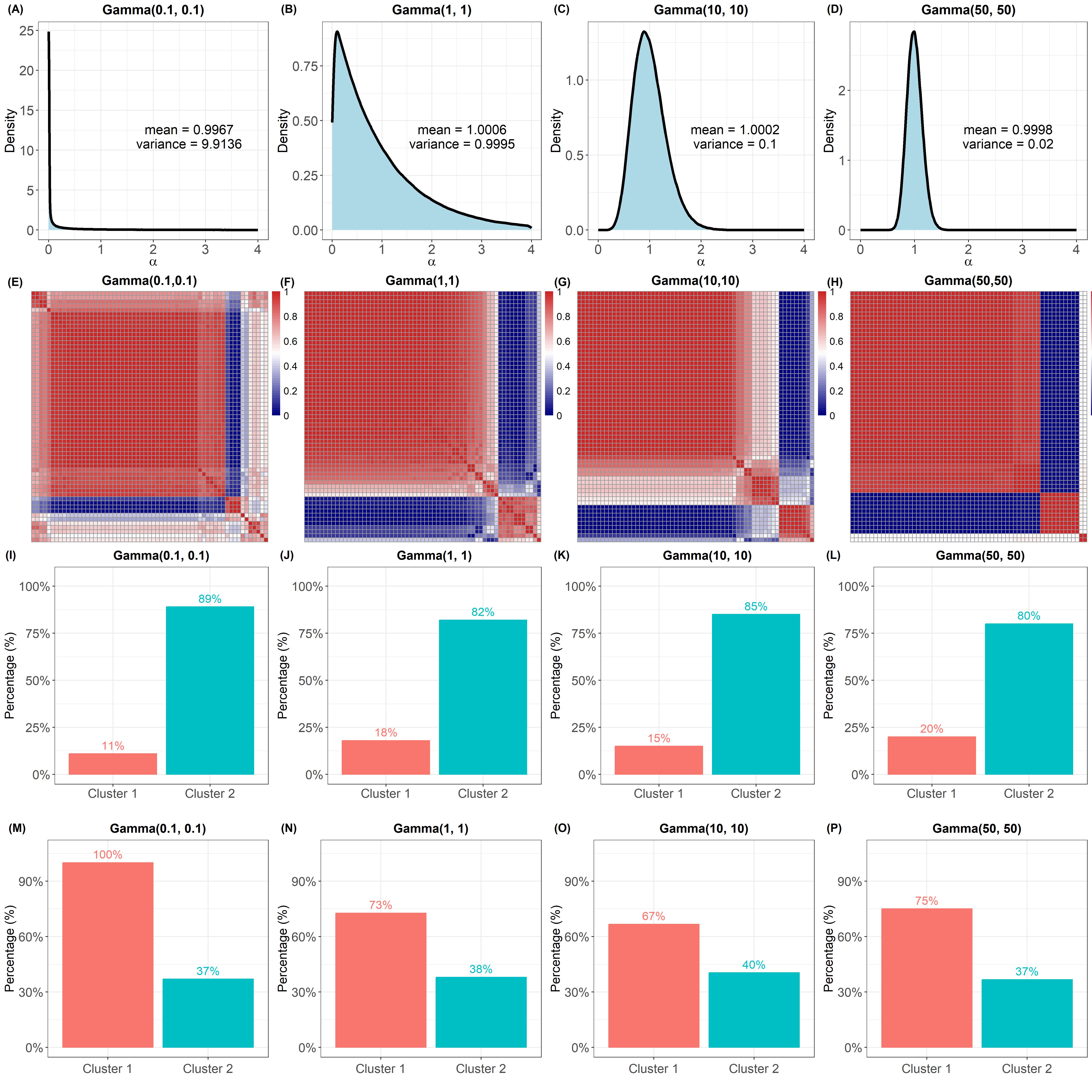}
\caption{Prior distributions for the concentration parameter $\alpha$, Posterior similarity matrix, the cluster proportions and the prevalence of insulin resistance by clusters for the longitudinal cytokines data (Example 1). 
(A) Prior distribution  for $\alpha \sim \text{Gamma} (0.1, 0.1) $. (B) Prior distribution for  $\alpha \sim \text{Gamma} (1, 1)$. 
(C) Prior distribution  for $\alpha \sim \text{Gamma} (10, 10)$. (D) Prior distribution  for  $\alpha \sim \text{Gamma} (50, 50)$.
(E) Posterior similarity matrix for $\alpha \sim \text{Gamma} (0.1, 0.1) $. (F) Posterior similarity matrix for $\alpha \sim \text{Gamma} (1, 1) $.
(G) Posterior similarity matrix for $\alpha \sim \text{Gamma} (10, 10) $. (H) Posterior similarity matrix for $\alpha \sim \text{Gamma} (50, 50) $.
(I)  Cluster proportions based on  $\alpha$ under prior  $\text{Gamma} (0.1, 0.1) $. (J) Cluster proportions based on  $\alpha$ under prior  $\text{Gamma} (1, 1) $. 
(K) Cluster proportions based on  $\alpha$ under prior  $\text{Gamma} (10, 10) $. (L) Cluster proportions based on  $\alpha$ under prior  $\text{Gamma} (50, 50) $.
(M) Proportion of insulin-sensitive individuals by clusters from the model with $\alpha \sim \text{Gamma} (0.1, 0.1) $. 
(N) Proportion of insulin-sensitive individuals by clusters from the model with  $\alpha \sim \text{Gamma} (1, 1) $. 
(O) Proportion of insulin-sensitive individuals by clusters from the model with $\alpha \sim \text{Gamma} (10, 10) $. 
(P)  Proportion of insulin-sensitive individuals by clusters from the model with  $\alpha \sim \text{Gamma} (50, 50) $.  }
\end{figure}

\begin{figure}[htbp]  
 \centering
   \includegraphics[width=13cm,height=7cm]{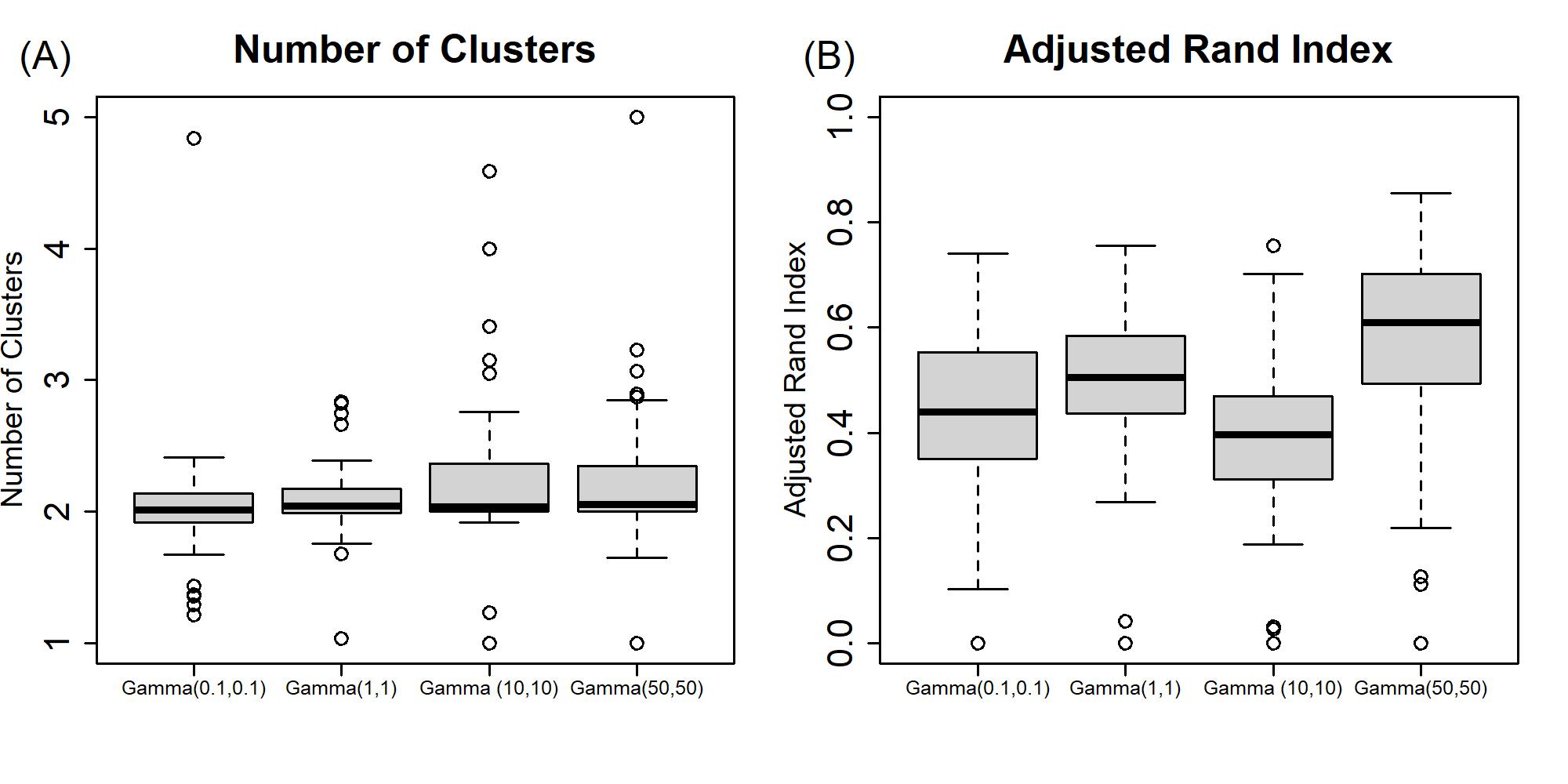}
\caption{Stability of the number of clusters and individual cluster membership over 61 data subsets under different prior for $\alpha$ for the longitudinal cytokines data (Example 1).
(A)  Distribution of the estimated number of clusters by prior distributions of $\alpha$. 
(B) Distribution of the adjusted Rand index by prior distributions of $\alpha$. }
\end{figure}

\subsection{Example 2: Primary Biliary Cholangitis Data}
\subsubsection{Data description}
Primary Biliary Cholangitis (PBC) is a chronic liver disease with an unknown cause. A randomized placebo-controlled trial of the drug D-penicillamine for PBC patients was conducted between 1974 and 1984 \citep{Dickson1989}.  During the study, multiple longitudinal markers were collected to monitor patients' disease progress, example variables include serum albumin, alkaline phosphotase, presence of ascites and serum bilirubin. The data are available at http://lib.stat.cmu.edu/datasets/pbcseq. Consistent with previous studies \citep{Komarek2013, Lu2022},  we only include $N = 260$ patients known to be alive at 910 days (30 months) of follow-up, and the longitudinal measurements up to this point. \par 

The goal of the current analysis was to identify patients' subgroups based on 10 longitudinal (continuous or categorical) biological markers, namely (1) presence of ascites (ascites), (2) presence of hepatomegaly or enlarged liver (hepatom), (3) the presence of blood vessel malformations in the skin (spiders), (4) presence of edema (edema.bin), (5) log of serum bilirunbin (mg/dl) (lbili), (6) log of serum albumin (g/dl) (lalbumin), (7) log of alkaline phosphotase (U/liter) (lalk.phos), (8) log of serum glutamic-oxaloacetic transaminase (lsgot), (9) log of platelet count (lplatelet), (10) log of standardized blood clotting time. Subsequently, to further characterize the resulting clusters, we associated the clusters with the survival status after 30 months using Kaplan-Meier analysis.  Time to death was considered as an event of interest and liver transplantation was considered as censoring. \par 

\subsubsection{Model specification}
We used the same specification hyper-parameters except we set $\nu_0 = 50 $ and $\bm{\Delta}$ to be a diagonal matrix with elements being 100. Similarly, to allow sufficient flexibility and to fully capture the dependence between markers and measurements within individuals, a linear form of random effects of time (i.e., both random intercept and slope) was used. This resulted in a dimension of 20 for random effect $\bm{\beta}_i$. In addition, the model adjusted for baseline age and sex.  We ran the model with 20000 iterations, discarded the first 10000 samples and kept every 10th sample. This resulted in 1000 samples for each model parameter.

To evaluate the cluster stability, cross-validation was used by creating 50 subsets from the original data by holding out 10 or 11 individuals. We refit the proposed model based on this subset and compared the results (both the estimated number of clusters and the individual cluster membership) to those obtained from the original data. Similarly, aRand was used to measure the agreement between the two clusterings.

\subsubsection{Analysis results}
For all models, the trace plots of MCMC samples suggested the model converged well (Figure S2). The clustering results based on the model under different prior distributions for $\alpha$ are shown in Figure S3.  In general, the number of clusters and the size of the clusters differ between models under different prior distributions for $\alpha$. Specifically, when $\alpha \sim \text{Gamma}(0.1, 0.1)$, two clusters were obtained (Figure S3A), with Cluster 1 accounting for 63.8\% of patients and Cluster 2 accounting for 36.2\% patients, respectively. Markers such as hepatom, spiders, edema.bin and lbili showed distinct patterns between the two clusters while others such as lalbumin, lalk.phos, lchol, lsogt, lplatelet and lprotime appeared to be similar between clusters (Figure S4). This suggested that patients belonging to Cluster 1 were more likely to have a presence of hepatomegaly or enlarged liver (hepatom), blood vessel malformations in the skin (spiders), presence of edema (edema.bin), and higher value of serum bilirunbin (mg/dl) (lbili). Therefore, Cluster 1 represented a group of patients with worse conditions and significantly lower survival probabilities after 30 months of follow-up, compared to Cluster 2 (Figure S3E). When $\alpha \sim \text{Gamma}(1, 1)$, three clusters were obtained (Figure S3B), with Cluster 1 accounting for 35.4\% of patients, Cluster 2 accounting for 26.5\% patients and Cluster 3 accounting for 38.1\% patients, respectively. The markers that were separated among clusters were also hepatom, spiders, edema.bin and lbili (Figure S5). Clusters 1, 2 and 3 represented patients that had mild, moderate and severe conditions reflected by their survival probabilities after 30 months (Figure S3F). However, the difference between Clusters 2 and 3 in survival probabilities was less evident. When $\alpha \sim \text{Gamma}(10, 10)$, three clusters were obtained (Figure S3C), with Cluster 1 accounting for 25.8\% of patients, Cluster 2 accounting for 33.5\% patients and Cluster 3 accounting for 40.8\% patients, respectively. The markers that were separated among clusters were also hepatom, spiders, edema.bin and lbili (Figure S6). Similarly, Clusters 1, 2 and 3 represented patients that had mild, moderate and severe conditions reflected by their survival probabilities after 30 months (Figure S3G). However, the difference between Clusters 1 and 2 in survival probabilities was minimal. When $\alpha \sim \text{Gamma}(50, 50)$, three clusters were obtained (Figure S3D), with Cluster 1 accounting for 35.4\% of patients, Cluster 2 accounting for 50.8\% patients and Cluster 3 accounting for 13.8\% patients, respectively. The markers that were separated among clusters were also hepatom, spiders, edema.bin and lbili (Figure S7). Similarly, Clusters 1, 2 and 3 represented patients that had mild, moderate and severe conditions reflected by their survival probabilities after 30 months (Figure S3H). However, the difference between Clusters 2 and 3 in survival probabilities was minimal. The stability of the number of clusters and individual cluster membership under different prior distributions for $\alpha$ over 50 random subsets is shown in Figure S8. A large variation was observed for the number of clusters and the aRand for the prior distribution of $\text{Gamma}(10, 10)$ and $\text{Gamma}(50, 50)$.

\section{Simulation Study}
In this section, we perform a simulation study to evaluate the clustering performance of the proposed model under various settings compared to several alternative approaches. 

\subsection{Simulation setup}
Data were generated according to equation (2) with the identity link function. Number of observations was set to be $n_{ir} = 6$ for $ i =1,..., N$ and $r = 1,...,R$. The individual visit time $t_{irj}$ for $ r  = 1,..., R$ and $j = 1,..., n_{ir}$ was sampled from a uniform distribution for the intervals of  (0, 2), (3, 5), (6, 8), (12, 14), (18, 20), (24, 26), (30, 32) respectively. The number of features that inform clustering structure was set to be $R_{\text{signal}} = 3$. For these three features, the random effects (both a random intercept and random slope) were generated from a joint multivariate normal distribution $\text{MVN}(\bm{0}, \bm{\Sigma}_\beta)$. The number of features that were noise (i.e., not informing the clustering structure) was denoted as $R_{\text{noise}}$ and was set to be  25, 50, 75, and 100, respectively. Therefore, the total number of features were $R = R_{\text{signal}}  + R_{\text{noise}}$.  A time-independent binary covariate was generated from a $\text{Bernouli}(0.5)$ distribution, with fixed effect coefficients set to be -0.5.  The number of clusters was set to be $K = 2, 3, 4 $, which is commonly seen in clinical practice. The sample size for each cluster was set to be $N_k = 50$, for $k = 1,..., K$. Two different scenarios were considered. In Scenario 1, the patterns of the features informing the clustering structure were relatively constant over time with a slope near zero (Figure S9) while in Scenario 2, a non-zero slope was added to the clusters leading to the trajectory pattern being increased or decreased over time (Figure S10).  

We compared the proposed model (called \textbf{lamb\_long}) to the following four approaches. 
\begin{itemize}
	\item   \textbf{lamb\_twostage}: in the first stage, a mixed-effect model was fitted for each longitudinal response and the estimates of the random intercepts and slopes were obtained for each individual. In the second stage, lamb model was applied to the joint random effects to obtain the optimal number of clusters and the cluster partition.
	\item   \textbf{lamb\_first}: applying lamb model to only the first measurement of each individual. 
	\item   \textbf{lamb\_last}: applying lamb model to only the last measurement of each individual. 
	\item   \textbf{K-means clustering}: K-means clustering for multivariate longitudinal data using the  \textit{kml3d} package \citep{Genolini2013}.   For this approach, the number of clusters was determined by the built-in Calinski and Harabatz criterion \citep{Genolini2013}. 
	\item  \textbf{hierarchical clustering}: hierarchical clustering for multivariate longitudinal data using the  \textit{clusterMLD} package \citep{Zhou2023}.   For this approach, the number of clusters was determined by the built-in Gap statistics \citep{Zhou2023}. 
\end{itemize}
For each setting, 50 datasets were generated. To measure the model performance, we calculated the mean and standard deviation (SD) of the estimated number of clusters ($\widehat{K}$) from each model over all the simulated datasets. To evaluate the model performance on recovering the true underlying cluster structure (i.e., partition), we used the aRand \citep{Hubert1985} to measure the agreement between the estimated cluster partition and the true cluster partition. To facilitate a fair comparison, the aRands for K-means and hierarchical clustering were calculated based on the true number of clusters $K$ (instead of the estimated one). 

\subsection{Simulation results}
The mean estimated number of clusters and aRand for Scenarios 1 and 2 are plotted in Figures 5 \& 6, respectively.  For Scenario 1, \textbf{lamb\_long}, \textbf{lamb\_first} and \textbf{lamb\_last} yielded a reasonably well estimate of the number of clusters for different $K$ (Figure 5A) as well as performing well in recovering the true clustering (Figure 5B). The performance of  \textbf{lamb\_first} and \textbf{lamb\_last} was expected given the clustering can be well determined by the first or last measurement of the individuals (Figure S9). On the other hand, \textbf{lamb\_twostage}, \textbf{K-means} \textbf{Hierachical clustering} underestimated or overestimated the number of clusters (Figure 5A) and failed to uncover the true underlying clustering structure particularly when the number of noise variable ($R_{noise}$) and $K$ were large.   The mean (SD) of the number of clusters and aRand under different $\alpha$ as well as the competing approaches for Scenario 1 are presented in Table S2.
 
\begin{figure}[H]  
 \centering
   \includegraphics[width=14cm,height=10cm]{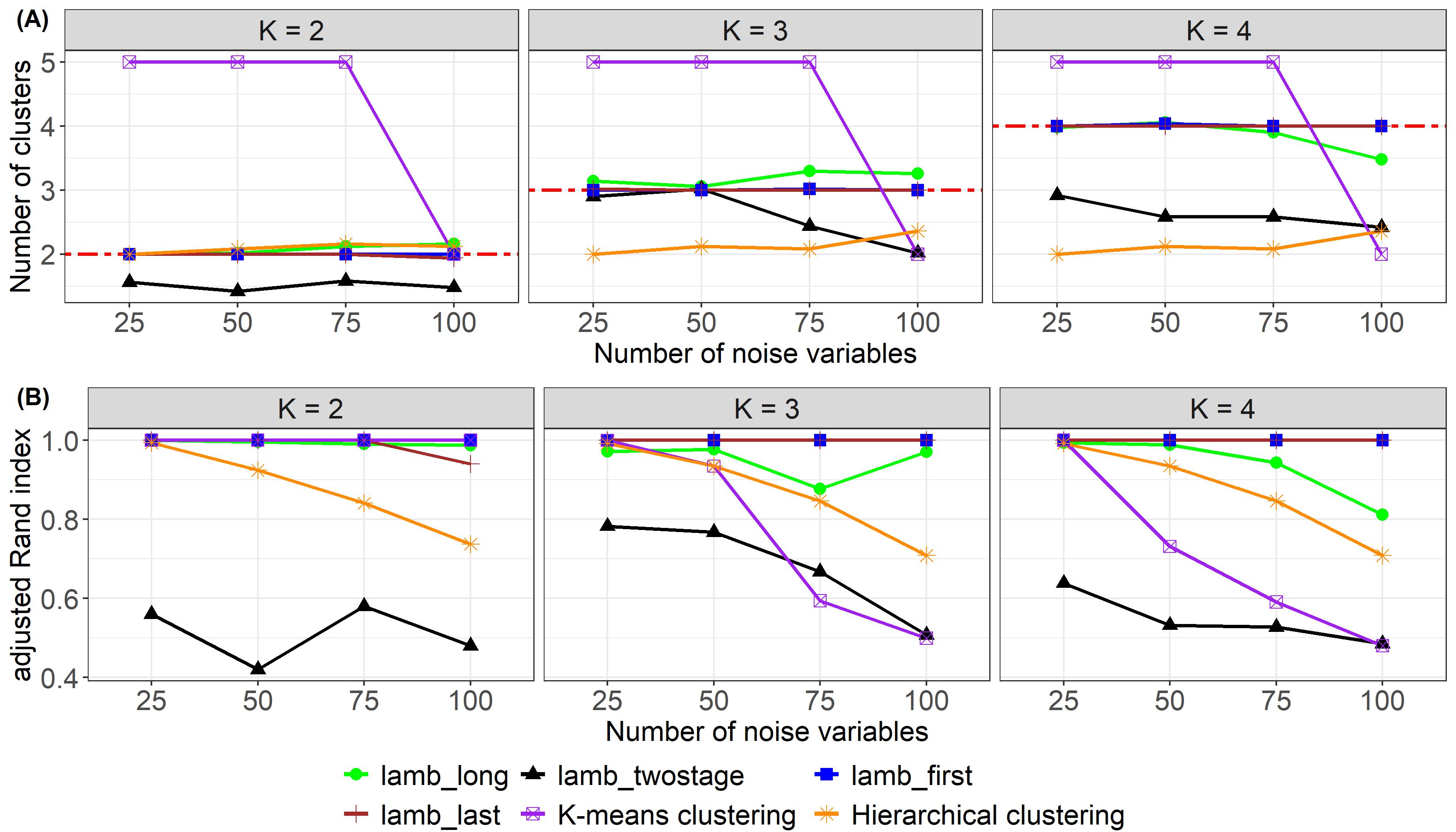}
\caption{Simulation results for Scenario 1. (A)  Mean estimated number of clusters over 50 simulated datasets by methods for $K =$ 2, 3 and 4. The reference line (in red) represents the true number of clusters. 
(B)  Mean adjusted rand index over 50 datasets by methods for $K =$ 2, 3 and 4.  
\textbf{lamb\_long}: lamb for longitudinal data with prior $\alpha \sim \text{Gamma}(0.1, 0.1)$. \textbf{lamb\_twostage}:  lamb for longitudinal data based on a two-stage approach. \textbf{lamb\_first}:  lamb based on the first measurement of each individual. \textbf{lamb\_first}:  lamb based on the last measurement of each individual.  \textbf{K-means}: K-means clustering for multivariate longitudinal data using kml3d package.  \textbf{Hierarchical clustering}: Hierarchical clustering for multivariate longitudinal data using clusterMLD package. For both K-means and Hierarchical clustering, the aRands were calculated under the true number of clusters $K$. }
\end{figure}

For Scenario 2, \textbf{lamb\_long} yielded a reasonably well estimate of the number of clusters for different $K$ (Figure 6A) and performed well in uncovering the true partition (Figure 6B), outperforming the competing approaches most of the time. On the other hand,  \textbf{lamb\_twostage}, \textbf{lamb\_first} and \textbf{lamb\_last} underestimated the number of clusters (Figure 6A), leading to poor performance in recovering the true partition (Figure 6B). The \textbf{K-means} and \textbf{Hierarchical} clustering approaches underestimated or overestimated the number of clusters (Figure 6A), particularly when the number of noise variables was large (e.g., $>50$). Given the true number of cluster $K$, the \textbf{K-means} approach performed reasonably well in recovering the true partition when $K=2$ (Figure 6B). However, the \textbf{K-means} approach yielded poor performance when $K = 3$ and $K = 4$, particularly when $R_{noise}$ is large (e.g., $\ge 50$). The mean (SD) of the number of clusters and aRand under different $\alpha$ as well as the competing approaches for Scenario 2 are presented in Table S3.

\begin{figure}[H]  
 \centering
   \includegraphics[width=14cm,height=10cm]{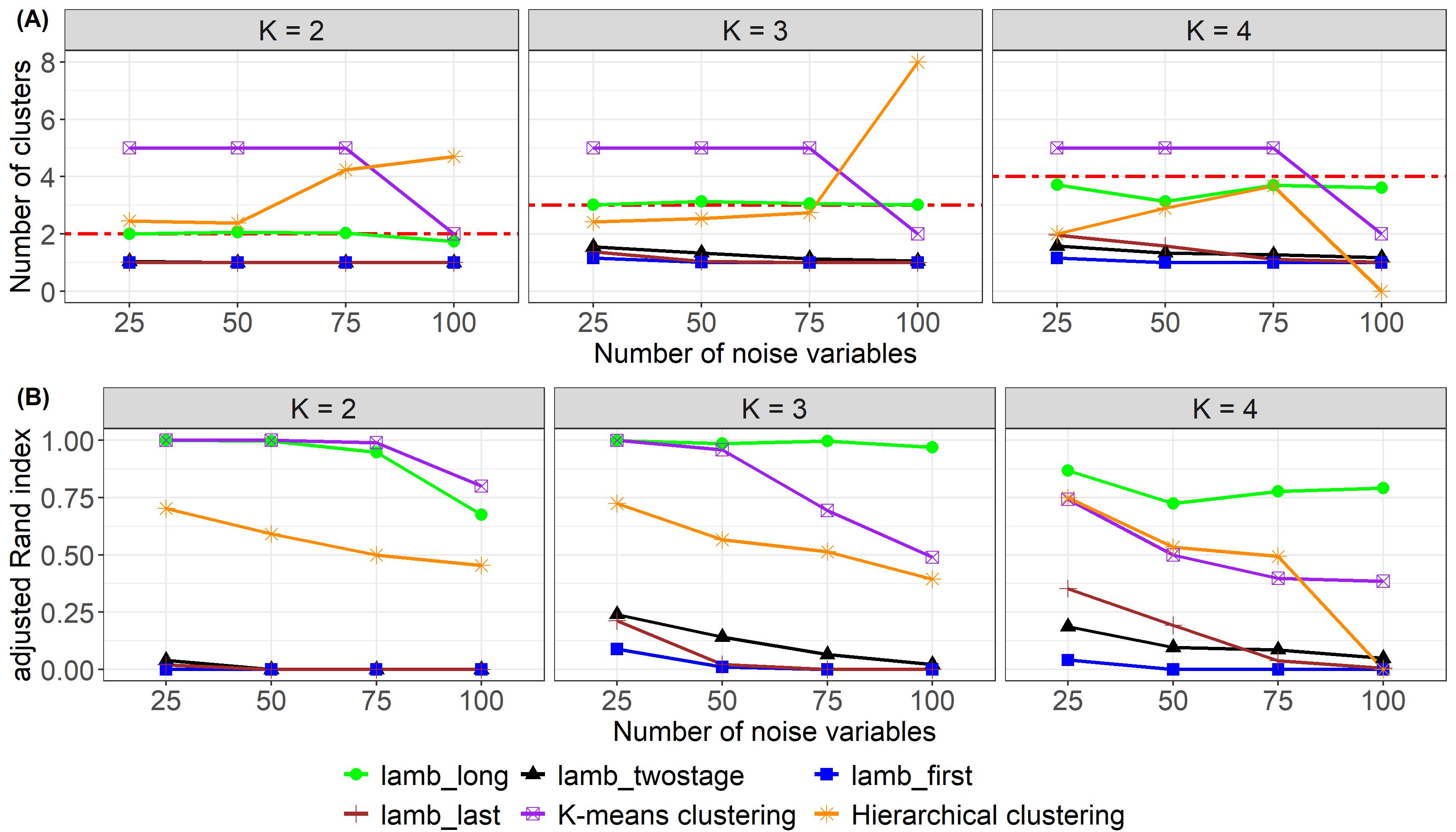}
\caption{Simulation results for Scenario 2. (A)  Mean estimated number of clusters over 50 simulated datasets by methods for $K =$ 2, 3 and 4. The reference line (in red) represents the true number of clusters. 
(B)  Mean adjusted rand index over 50 datasets by methods for $K =$ 2, 3 and 4. 
\textbf{lamb\_long}: lamb for longitudinal data with prior $\alpha \sim \text{Gamma}(0.1, 0.1)$. \textbf{lamb\_twostage}:  lamb for longitudinal data based on a two-stage approach. \textbf{lamb\_first}:  lamb based on the first measurement of each individual. \textbf{lamb\_first}:  lamb based on the last measurement of each individual. \textbf{K-means}: K-means clustering for multivariate longitudinal data using kml3d package.  \textbf{Hierarchical clustering}: Hierarchical clustering for multivariate longitudinal data using clusterMLD package. For both K-means and Hierarchical clustering, the aRands were calculated under the true number of clusters $K$.  }
\end{figure}

\section{Discussion}
In the current study, we propose a Bayesian nonparametric model for clustering high-dimensional longitudinal data. The underlying assumptions are that the longitudinal data are correlated within and between features and their dependence can be captured by jointly modeling the random effects, which can be further represented by low-dimensional latent factors. 

The proposed model provides a new perspective and an analytical tool for clustering multi-dimensional longitudinal features, as demonstrated through our practical applications and simulation studies. In the cytokines data example, the underlying heterogeneity captured by the proposed model reflects a distinct prevalence of insulin resistance, which suggests a biological difference between the resulting clusters. In the PBC data example, the resulting clusters differed on certain markers but not others, suggesting that not all markers contributed to defining the clusters and it is essential to perform the clustering on a lower dimensional space in which the noise information was removed. This is particularly useful in practice because which features will contribute to the clusterings can be hardly known a priori. Reducing noise and performing the analysis on a lower dimensional space could lead to a more stable and clinically meaningful clustering result. The simulation study demonstrates that the proposed model provides a reasonable well estimation of the number of clusters and the cluster membership under various scenarios with different numbers of clusters and noise features. \par

There are several directions for future studies. For example, the current model is computed using the Gibbs sampling and split-merge algorithm. To scale up the algorithm to analyze large datasets (e.g., electronic health record data), variational inference can be considered as an alternative \citep{blei2006}.   In addition, the trajectory can be analyzed through non-parametric functions such as splines to allow for more flexibility in capturing non-linear forms. Finally, our model reduces the dimension of features through latent factors. However, in some clinical settings, determining which variables contribute to the final clustering is of interest, and therefore simultaneously clustering and variable selection \citep{Kim2006} can be considered in future studies. \par

Nevertheless, the proposed model serves as a useful tool for clustering high-dimensional data with complex data structures and an unknown number of clusters. 

\section*{Data Availability Statement}
The longitudinal cytokine data are available from \citet{Sailani2020} and are hosted on the NIH Human Microbiome 2 project site (https://portal.hmpdacc.org). The PBC data are available from Fleming and Harrington, Appendix D \citep{Fleming2013} or in R mixAK package \citep{Komarek2014}.

\section*{Funding}
Z.L. is supported by a Discovery Grant funded by the Natural Sciences and Engineering Research Council of Canada. 


\end{document}


\maketitle

\setcounter{figure}{0}
\renewcommand\thefigure{S\arabic{figure}}
\setcounter{table}{0}
\renewcommand\thetable{S\arabic{table}}

\section{Posterior computation}
To obtain the posterior distribution of model parameters for the proposed model, we developed a Gibbs sampler. Let $|\cdot$ denote that the distribution is conditional on all other parameters. The posterior requires the full conditional distribution of each parameter. The Gibbs sampler cycles through the following steps: 
\begin{itemize}
	\item  Update $q\times d$ factor loadings matrix $\bm{\Lambda}$: sample $\bm{\lambda}_k$, $\varphi$, $\tau$ and $\phi$ from the following posteriors:
	\begin{enumerate}
	\item If we denote the $k$th row of $\bm{\Lambda}$ by $\bm{\lambda}_k^\top$, for $k=1,...,q$, where $q$ is the dimension of the joint random effect $\bm{\beta}_i$, then the $\bm{\lambda}_k s$ have independent conditionally conjugate posteriors:
		$$ \bm{\lambda}_k |\cdot \sim \text{MVN}_{d}( ( \bm{D}^{-1}_k + \omega^{-2}_{\beta k}  \bm{\eta}^\top \bm{\eta} )^{-1}\bm{\eta}^\top \omega_{\beta k} ^{-2} \bm{\beta}^{(k)}, ( \bm{D}^{-1}_k + \omega^{-2}_{\beta k} \bm{\eta}^\top \bm{\eta} )^{-1}  ) $$
	where  $ \bm{D}_k = \tau^2 \text{diag}(\varphi_{k1}\phi^2_{k1},...,\varphi_{kd}\phi^2_{kd}) $, $\bm{\eta} = (\bm{\eta}_1,...,\bm{\eta}_N)^\top$ and $\bm{\beta}^{(k)}  = (\beta_{1k},..., \beta_{Nk})^\top$ denoting the $k$th column of the random effect matrix, for $k=1,...,q$, . 
	\item Update $\varphi_{kh}$. First sample $\tilde{\varphi}_{kh}$, for $k=1,...,q$ and $h=1,...,d$ from an inverse-Gaussian $\text{iG}(\tau\phi_{kh}/|\lambda_{kh}|, 1) $ distribution and set $ \varphi_{kh} = 1/ \tilde{\varphi}_{kh}$.
	\item Update $\tau$. The posterior of $\tau$ is a generalized inverse Gaussian distribution. 
	$$\tau | \cdot   \sim \text{giG}\{dq(1-a), 1, 2 \sum_{k,h}|\lambda_{kh}|\phi_{kh} \} $$
	\item  Update $\phi_{kh}$. Draw $T_{kh}$ independently with 
		$T_{kh} | \cdot \sim \text{giG}(a-1, 1, 2|\lambda_{kh}|)$ and set $\phi_{kh} = T_{kh}/T$ with $T = \sum_{kh}T_{kh}$.
	\end{enumerate}	
 
\item Update $\omega^2_{\beta k}$, for $k=1,..,q$, the diagonal elements of $\bm{\Sigma}_\beta$.
			$$ \omega^2_{\beta k} | \cdot \sim \text{IG}(a_\omega + \frac{N}{2}, b_\omega + \sum_{i=1}^N \frac{(\bm{\beta}_{i} - \bm{\Lambda}\bm{\eta}_i)^2}{2})  $$

\item  Update  the $\bm{\Delta}_h$, for $h=1,...,\infty$. The $\bm{\Delta}_h$ is sampled from the inverse-Wishart distribution 
	$$ \bm{\Delta}_h | \cdot \sim \text{IW}(\hat{v}_h,\hat{\varphi}_h)$$
    where 
    $$ \hat{\varphi}_h = v_0 + n_h$$
     $$ \hat{\varphi}_h = \xi^2 \bm{I}_d + \sum_{i:c_i=h}(\bm{\eta}_i - \bar{\bm{\eta}}_h)(\bm{\eta}_i - \bar{\bm{\eta}}_h)^\top + \frac{\kappa_0n_h}{\kappa_0 + n_h} \bar{\bm{\eta}}_h\bar{\bm{\eta}}_h^\top \quad \text{and} \quad \bar{\bm{\eta}}_h = \frac{1}{n_h} \sum_{i:c_i=h}\bm{\eta}_i $$

    \item  Update the latent factors $\bm{\eta}_i$, for $ i = 1,...,N$. The  $\bm{\eta}_i$ can be sampled from 
			$$ \bm{\eta}_i | \cdot \sim \text{MVN}_d( \bm{\Omega}_h \rho_h, \bm{\Omega}_h + \bm{\Omega}_h (\hat{\kappa}_{h,-i}\bm{\Delta}_h)^{-1}\bm{\Omega}_h) $$  
      where $\hat{\kappa}_{h,-i} = \kappa_0 + N_{h, -i}$ with $N_{h, -i} = \sum_{j \neq i}\mathds{1}(c_j = h)$, $\bar{\eta}_{h,-i} = \frac{1}{N_{h,-i}}\sum_{j:c_j=h, j \neq i} \bm{\eta}_i$, $\hat{\mu}_{h,-i} = \frac{N_{h,-i}\bar{\eta}_{h,-i}}{N_{h,-i} + \kappa_0} $, $\rho_h = \bm{\Lambda}^\top \bm{\Sigma}^{-1}_\beta \bm{\beta}_i + \bm{\Delta}^{-1}_h \hat{\mu}_{h, -i}$ and $\bm{\Omega}^{-1}_h = \bm{\Lambda}^\top \bm{\Sigma}_\beta \bm{\Lambda} + \bm{\Delta}^{-1}_h$.

 \item Update the cluster indicator $c_i$ for $i = 1,...,n$. 
   \begin{equation}
   P(c_i = h |\cdot) \propto  
    \begin{cases}
      N_{h, -i} & \int \text{MVN}_d(\bm{\eta}_i; \bm{\mu}_h, \bm{\Delta}_h) d f( \bm{\mu}_h, \bm{\Delta}_h|c_{-i}, \bm{\eta}_{-i}) \quad \text{for} \quad h\in c_{-i} \\
     \alpha & \int \text{MVN}_d(\bm{\eta}_i; \bm{\mu}_h, \bm{\Delta}_h) d f( \bm{\mu}_h, \bm{\Delta}_h) \quad \text{for} \quad h \notin c_{-i}\\
    \end{cases}       
\end{equation}  
where $\bm{\eta}_{-i} = \{\bm{\eta}_j: j \neq i \}$ and $c_{-i} = \{c_j: j \neq i \}$.  
	\item Update concentration parameter $\alpha$.  Let $r$ be the number of unique $c_i's$ (number of clusters). Generate 
	$$ 
	\alpha | \cdot  \propto  
    \begin{cases}
      \text{Gamma}(\alpha + r, b_\sigma - \text{log} \varphi ) \quad \text{with probability } \pi \\
      \text{Gamma}(\alpha + r - 1, b_\sigma - \text{log} \varphi ) \quad \text{with probability } 1-\pi \\
    \end{cases}    
   $$

	\item Update random effect coefficients  $\bm{\beta}_{i} = (\bm{\beta}_{i1},..., \bm{\beta}_{iR})^\top$, for $i=1,..., N$, using Metropolis-Hastings algorithm. The update is based on the following full conditional distribution of the random effect
$$f(\bm{\beta}_i) \propto \text{exp}\Big\{  \sum_{r=1}^R \vartheta_r^{-1}(\bm{y}_{ir}^\top \bm{x}_{ir} - \bm{1}^\top \bm{Q}_{ir} ) - \frac{1}{2}  (\bm{\beta}_i - \bm{\Lambda}\bm{\eta}_i)^\top \bm{\Sigma}_{\beta}^{-1}(\bm{\beta}_i - \bm{\Lambda}\bm{\eta}_i) \Big\} $$
The detailed updating step can be found in previous studies \citep{Komarek2013, Tan2022a}.  In the case when all longitudinal features are continuous, the $\bm{\beta}_{i}$ can be sampled from 
	$\bm{\beta}_{i}|\cdot \sim \text{MVN}(\tilde{\bm{\mu}}_\beta, \tilde{\bm{\Sigma}}_\beta)  $, where  $\tilde{\bm{\Sigma}}_\beta =  \bigg({\bm{\Sigma}^{-1}_\beta + \text{block-diag} (\sigma_{r}^{-2}  \bm{Z}_{ir}^\top\bm{Z}_{ir}}: r=1,...,R \big) \bigg)^{-1}$ and  
	$\tilde{\bm{\mu}}_\beta = \tilde{\bm{\Sigma}}_\beta \bigg( 
	 \left(\begin{smallmatrix}
		\sigma^{-2}_1  \bm{Z}_{i1}^{\top} \big(\bm{y}_{i1} - \bm{x}_{i1}^{\top}\bm{\gamma}_{1} \big) \\
       	 \vdots \\
		\sigma^{-2}_{R} \bm{Z}_{iR}^{\top} \big(\bm{y}_{iR} - \bm{x}_{iR}^{\top}\bm{\gamma}_{R} \big) \\   
		\end{smallmatrix} \right) + \bm{\Sigma}_\beta^{-1} \bm{\Lambda} \bm{\eta}_i \bigg)$.

		\item  Update fixed effect coefficients $\bm{\gamma}_{r}|\cdot \sim \text{MVN}(\tilde{\bm{v}}_{r},\tilde{\bm{V}}_{r})$, where $\tilde{\bm{V}}_{r}=({\bm{V}^{-1}_{0r} + \frac{1}{\vartheta_{r}}\sum_{i=1}^N \bm{x}_{ir}^\top \bm{x}_{ir}})^{-1}$ and  $\tilde{\bm{v}}_{r}^{(s)} = \tilde{\bm{V}}_{r}(\frac{1}{\vartheta_{r}} \sum_{i=1}^N \bm{x}_{ir}^\top (\bm{y}_{ir} - \bm{Z}_{ir}^{\top} \bm{\beta}_{ir}))$.

	\item Update dispersion parameter $\vartheta_r$ for features that are normally distributed. In such a case, $\vartheta_r = \sigma_r^2$ and $\sigma^{2}_{r}|\cdot \sim \text{IG}(\tilde{a}_{r},\tilde{b}_{r})$, where $\tilde{a}_{r} = a_{\sigma} + \frac{1}{2}\sum_{i=1}^N n_{ir} $ and  $\tilde{b}_{r} = b_{\sigma} + \frac{1}{2}\sum_{i=1}^N||\bm{y}_{ir} - \bm{x}_{ir}^{\top}\bm{\gamma}_{r} - \bm{Z}^{\top}_{ir}\bm{\beta}_{ir}||^2$.
		\item Update the hyper-parameters of the Dirichlet-Laplace prior:
		\begin{itemize}
			\item Sample $\tilde{\varphi}_{jh}$, for $j=1,...,p$ and $h=1,...,d$ from an inverse-Gaussian $\text{iG}(\tau\phi_{jh}/|\lambda_{jh}|, 1) $ distribution and set $ \varphi_{jh} = 1/ \tilde{\varphi}_{jh}$.
			\item Sample $\tau$. The posterior of $\tau$ is a generalized inverse Gaussian distribution: 
	$\tau  \sim \text{giG}\{dp(1-a), 1, 2 \sum_{j,h}|\lambda_{jh}|\phi_{jh} \} $
			\item  Sample $\phi|\bm{\Lambda}$. Draw $T_{jh}$ independently with 
		$T_{jh} \sim \text{giG}(a-1, 1, 2|\lambda_{jh}|)$ and set $\phi_{jh} = T_{jh}/T$ with $T = \sum_{jh}T_{jh}$.
	\end{itemize}	
\end{itemize}

It is possible the simple Gibbs sampler is stuck in local modes leading to incorrect clustering or partition. Therefore, we employed a split-merge MCMC algorithm proposed by \citet{Jain2004}, which allows the Gibbs sampler to escape local modes when updating the cluster membership $c_i$ in formula (1). 


\clearpage
\newpage
\section{Supplementary Figures and Tables for the Real Data Application}
\begin{figure}[!htbp]  
 \centering
   \includegraphics[width=15cm,height=12cm]{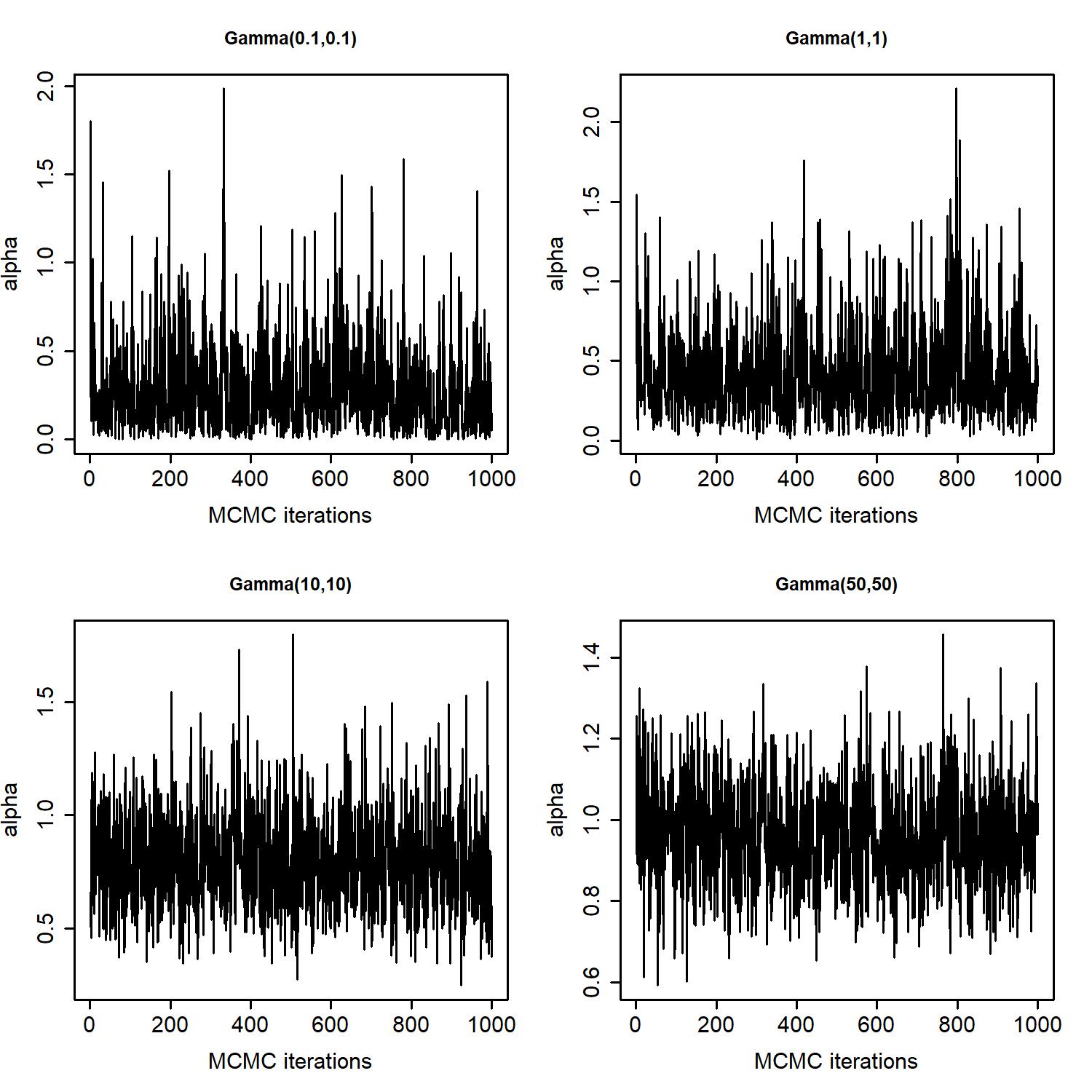}
\caption{Trace plot for concentration parameter $\alpha$ under different prior distributions for the longitudinal cytokines data  (Example 1). }
\end{figure}

\newpage
\begin{figure}[!htbp]  
 \centering
   \includegraphics[width=15cm,height=12cm]{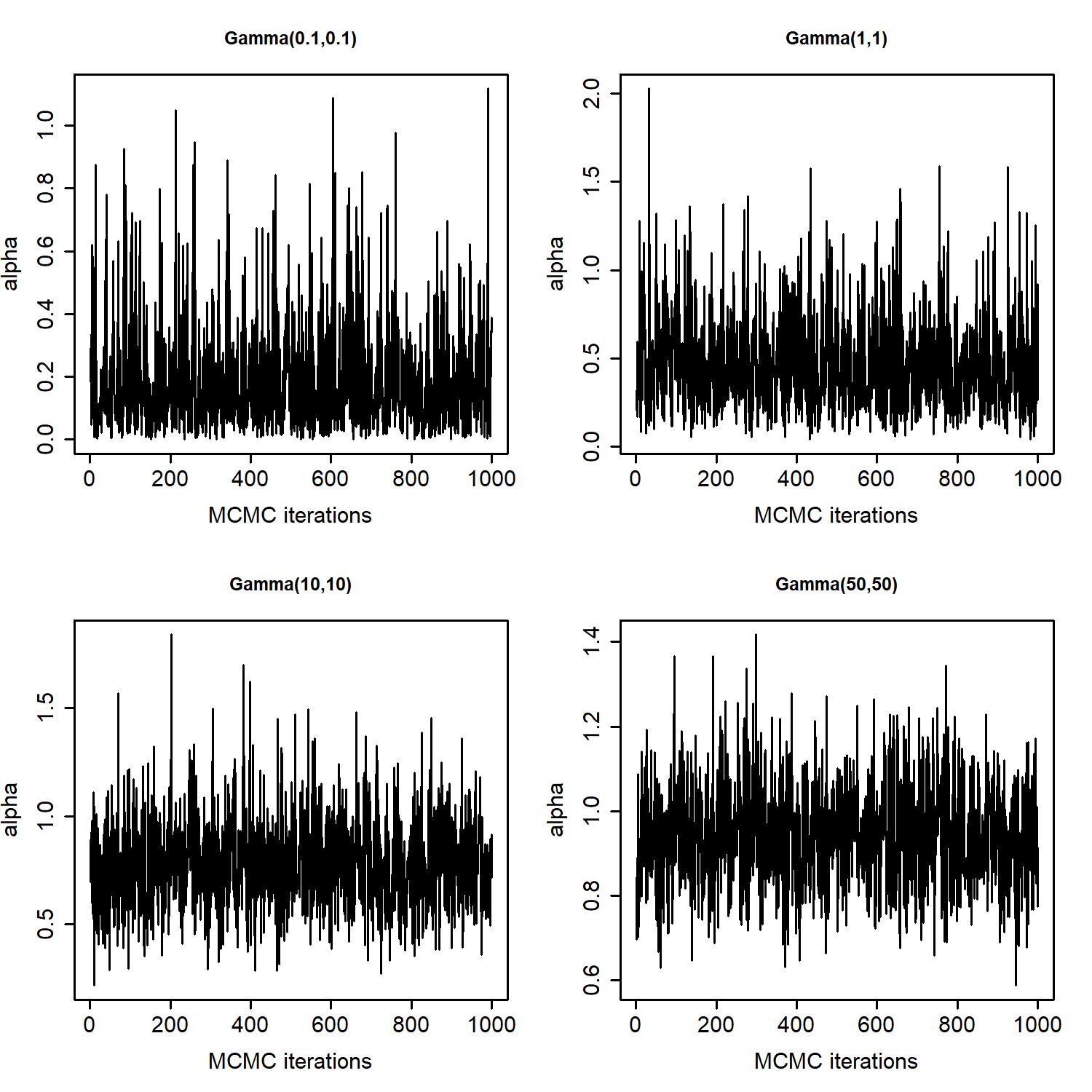}
\caption{Trace plot for concentration parameter $\alpha$ under different prior distributions for the PBC data  (Example 2). }
\end{figure}

\newpage
\begin{figure}[!htbp]  
 \centering
   \includegraphics[width=15cm,height=11cm]{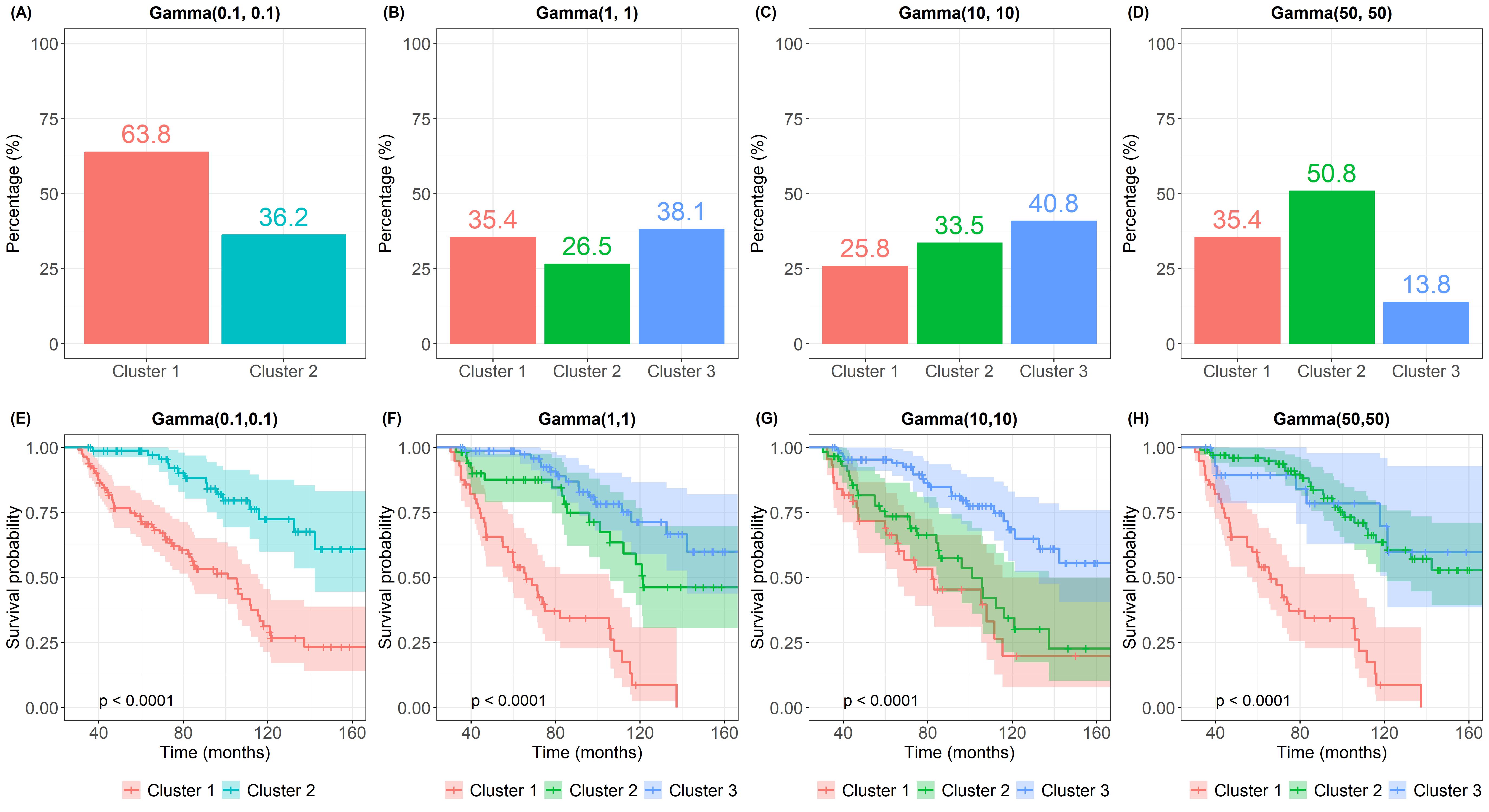}
\caption{Cluster proportions and survival probabilities under different prior distributions of $\alpha$ by clusters for the PBC data (Example 2). }
\end{figure}

\newpage
\begin{figure}[htbp]  
 \centering
   \includegraphics[width=14cm,height=10cm]{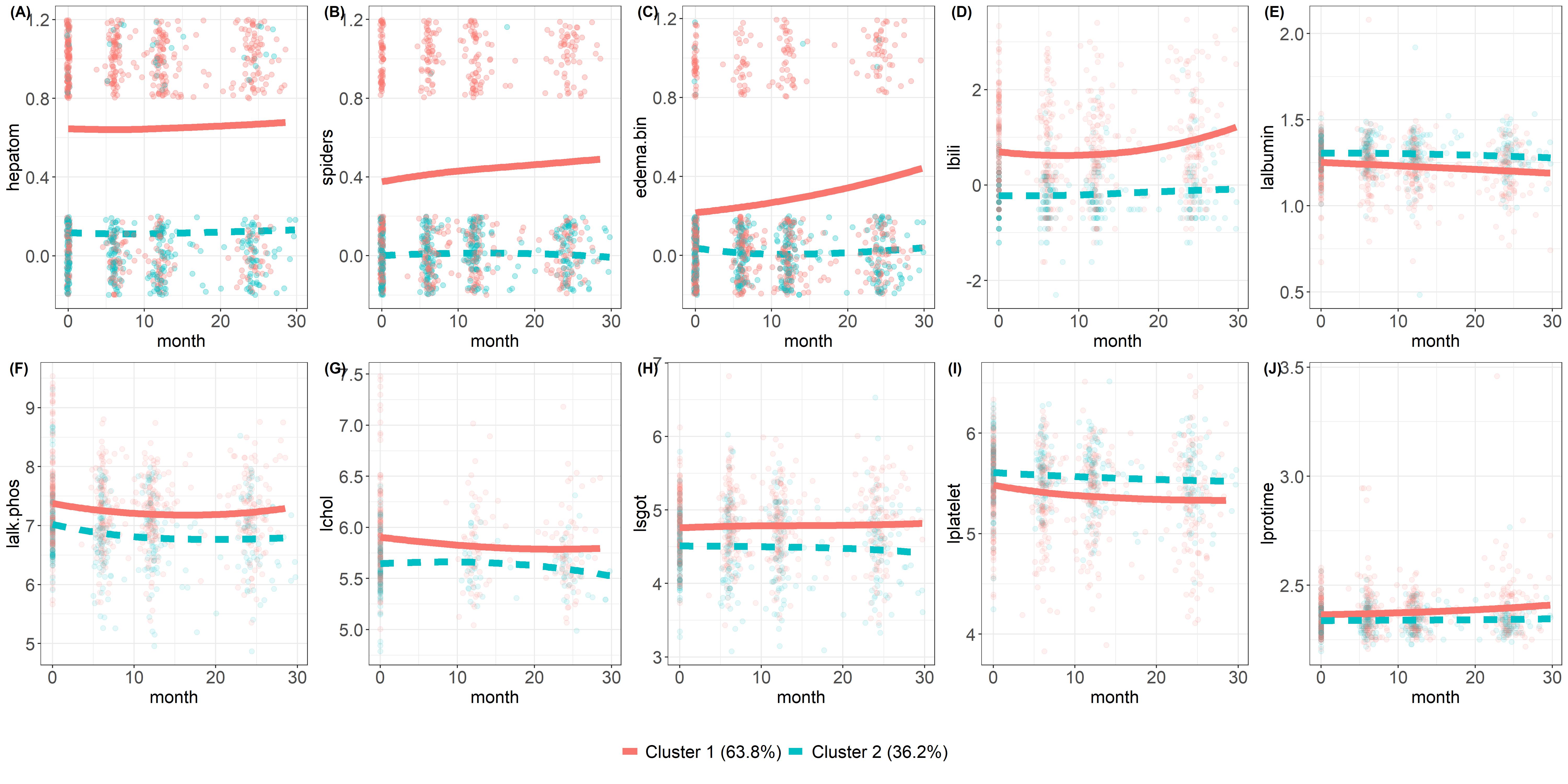}
\caption{Trajectory pattern by clusters for each longitudinal marker for the model under $\alpha \sim \text{Gamma}(0.1, 0.1)$ for the PBC data (Example 2). (A) Trajectory pattern for the presence of ascites (ascites). (B) Trajectory pattern for the presence of hepatomegaly or enlarged liver (hepatom), (C) Trajectory pattern for the presence of blood vessel malformations in the skin (spiders). (D) Trajectory pattern for the presence of edema (edema.bin). (E) Trajectory pattern for the  log of serum bilirunbin (mg/dl) (lbili). (F) Trajectory pattern for the log of serum albumin (g/dl) (lalbumin). (G) Trajectory pattern for the  log of alkaline phosphotase (U/liter) (lalk.phos). (H) Trajectory pattern for the log of serum glutamic-oxaloacetic transaminase (lsgot). (I) Trajectory pattern for the log of platelet count (lplatelet). (J) Trajectory pattern for the log of standardized blood clotting time.}
\end{figure}

\newpage
\begin{figure}[htbp]  
 \centering
   \includegraphics[width=14cm,height=10cm]{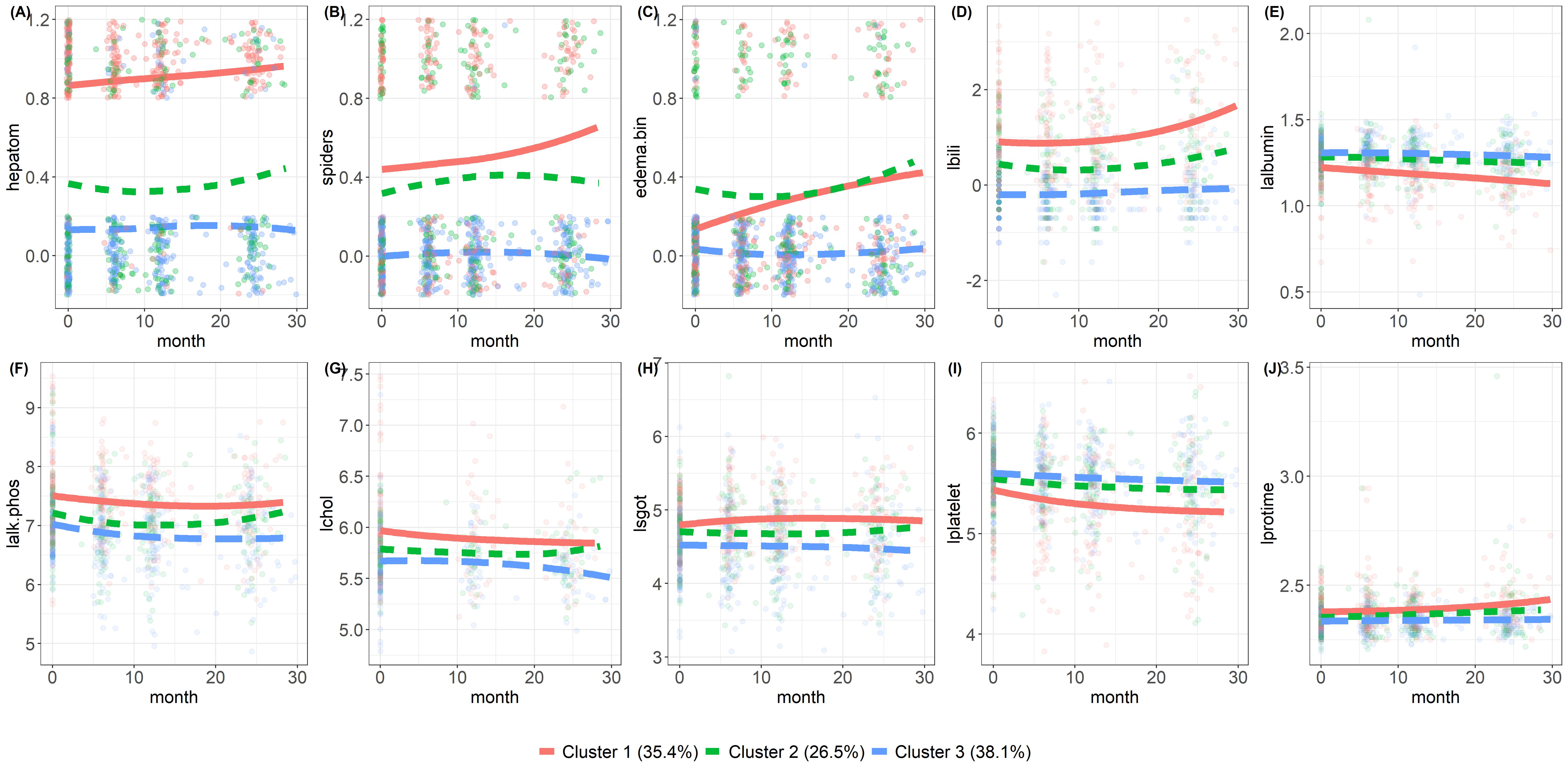}
\caption{Trajectory pattern by clusters for each longitudinal marker for the model under $\alpha \sim \text{Gamma}(1, 1)$ for the PBC data (Example 2). (A) Trajectory pattern for the presence of ascites (ascites). (B) Trajectory pattern for the presence of hepatomegaly or enlarged liver (hepatom), (C) Trajectory pattern for the presence of blood vessel malformations in the skin (spiders). (D) Trajectory pattern for the presence of edema (edema.bin). (E) Trajectory pattern for the  log of serum bilirunbin (mg/dl) (lbili). (F) Trajectory pattern for the log of serum albumin (g/dl) (lalbumin). (G) Trajectory pattern for the  log of alkaline phosphotase (U/liter) (lalk.phos). (H) Trajectory pattern for the log of serum glutamic-oxaloacetic transaminase (lsgot). (I) Trajectory pattern for the log of platelet count (lplatelet). (J) Trajectory pattern for the log of standardized blood clotting time.}
\end{figure}

\newpage
\begin{figure}[htbp]  
 \centering
   \includegraphics[width=14cm,height=10cm]{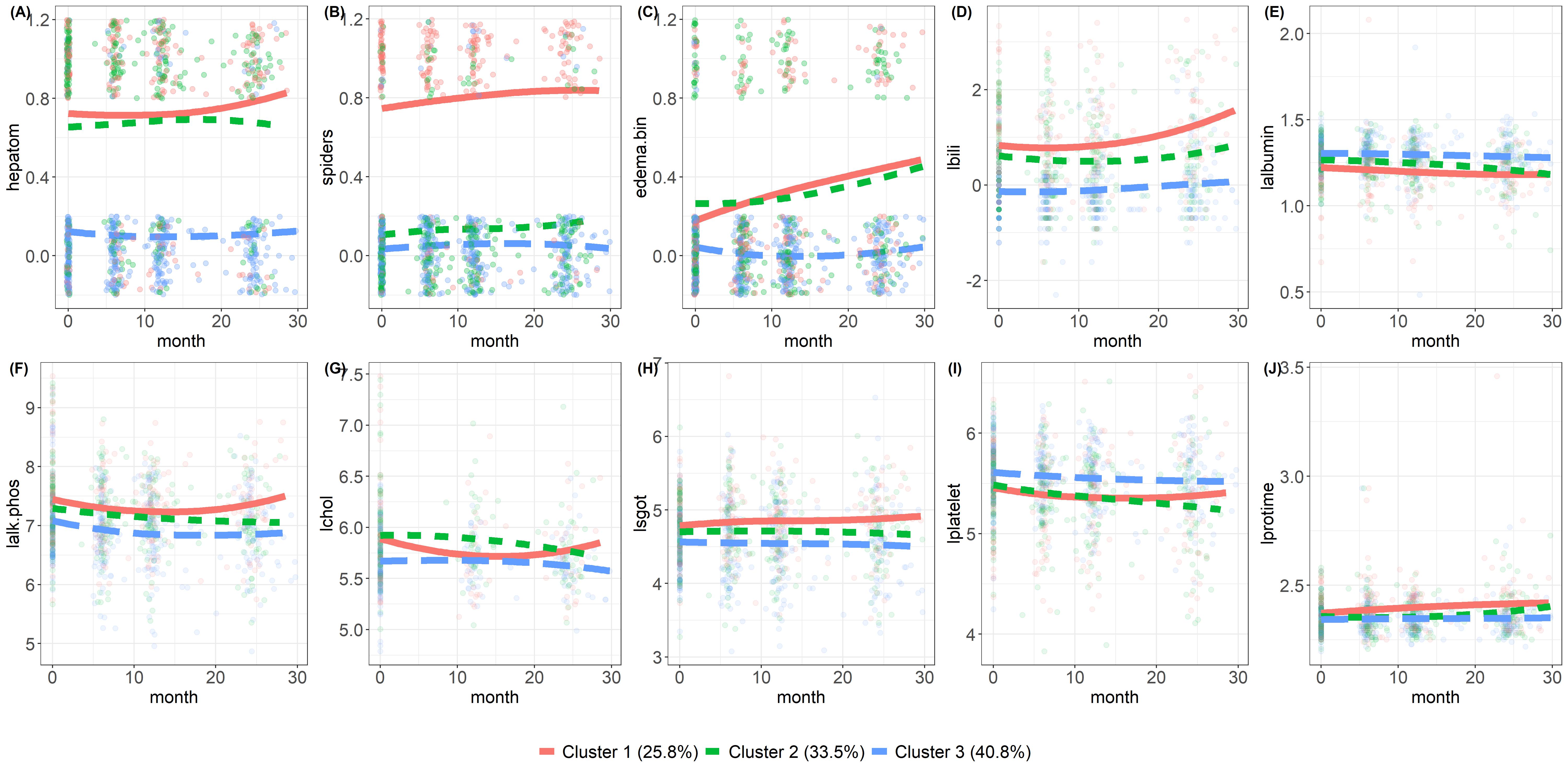}
\caption{Trajectory pattern by clusters for each longitudinal marker for the model under $\alpha \sim \text{Gamma}(10, 10)$ for the PBC data (Example 2). (A) Trajectory pattern for the presence of ascites (ascites). (B) Trajectory pattern for the presence of hepatomegaly or enlarged liver (hepatom), (C) Trajectory pattern for the presence of blood vessel malformations in the skin (spiders). (D) Trajectory pattern for the presence of edema (edema.bin). (E) Trajectory pattern for the  log of serum bilirunbin (mg/dl) (lbili). (F) Trajectory pattern for the log of serum albumin (g/dl) (lalbumin). (G) Trajectory pattern for the  log of alkaline phosphotase (U/liter) (lalk.phos). (H) Trajectory pattern for the log of serum glutamic-oxaloacetic transaminase (lsgot). (I) Trajectory pattern for the log of platelet count (lplatelet). (J) Trajectory pattern for the log of standardized blood clotting time.}
\end{figure}

\newpage
\begin{figure}[htbp]  
 \centering
   \includegraphics[width=14cm,height=10cm]{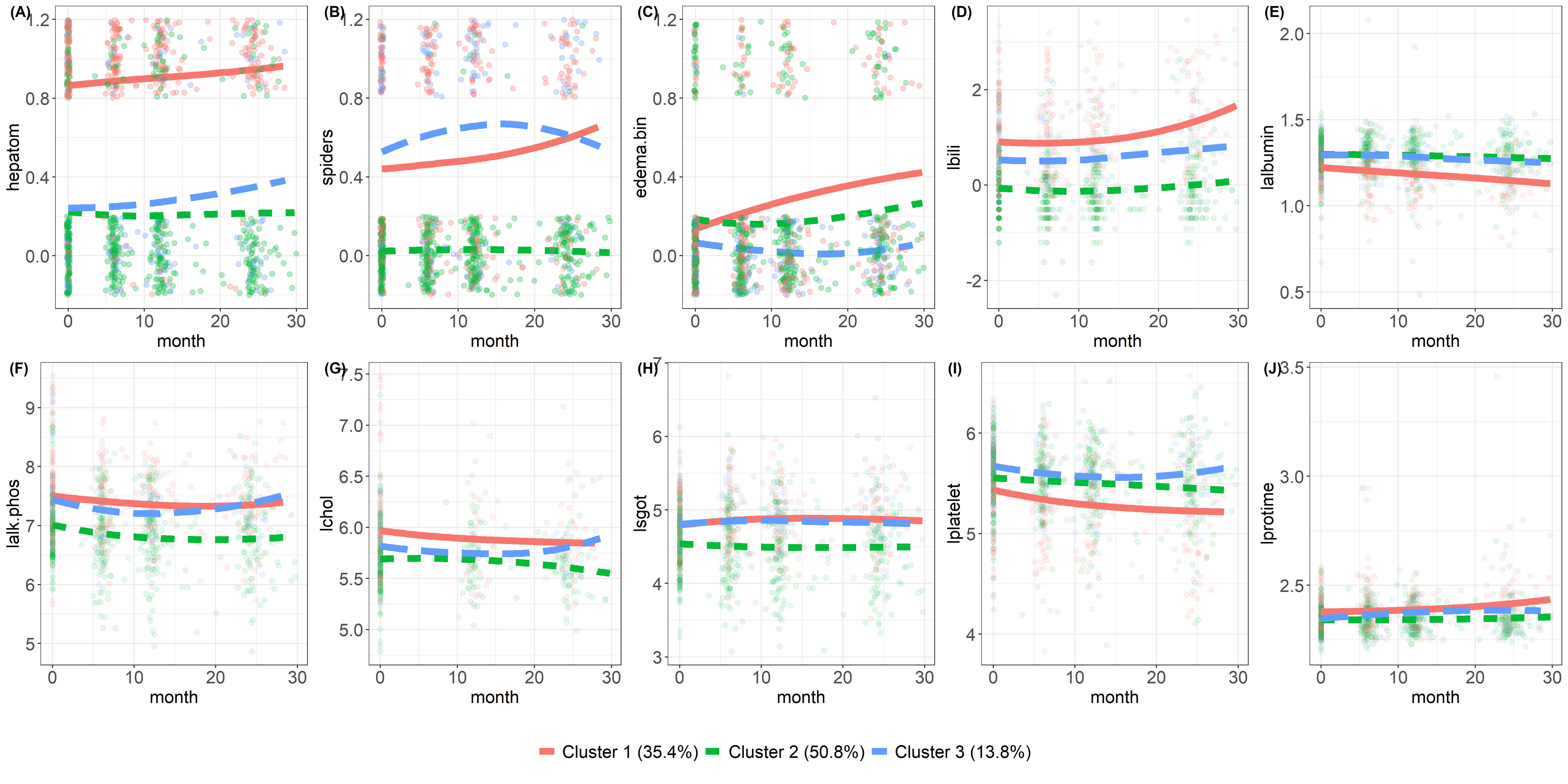}
\caption{Trajectory pattern by clusters for each longitudinal marker for the model under $\alpha \sim \text{Gamma}(50, 50)$ for the PBC data (Example 2). (A) Trajectory pattern for the presence of ascites (ascites). (B) Trajectory pattern for the presence of hepatomegaly or enlarged liver (hepatom), (C) Trajectory pattern for the presence of blood vessel malformations in the skin (spiders). (D) Trajectory pattern for the presence of edema (edema.bin). (E) Trajectory pattern for the  log of serum bilirunbin (mg/dl) (lbili). (F) Trajectory pattern for the log of serum albumin (g/dl) (lalbumin). (G) Trajectory pattern for the  log of alkaline phosphotase (U/liter) (lalk.phos). (H) Trajectory pattern for the log of serum glutamic-oxaloacetic transaminase (lsgot). (I) Trajectory pattern for the log of platelet count (lplatelet). (J) Trajectory pattern for the log of standardized blood clotting time.}
\end{figure}

\newpage
\begin{figure}[htbp]  
 \centering
   \includegraphics[width=13cm,height=7cm]{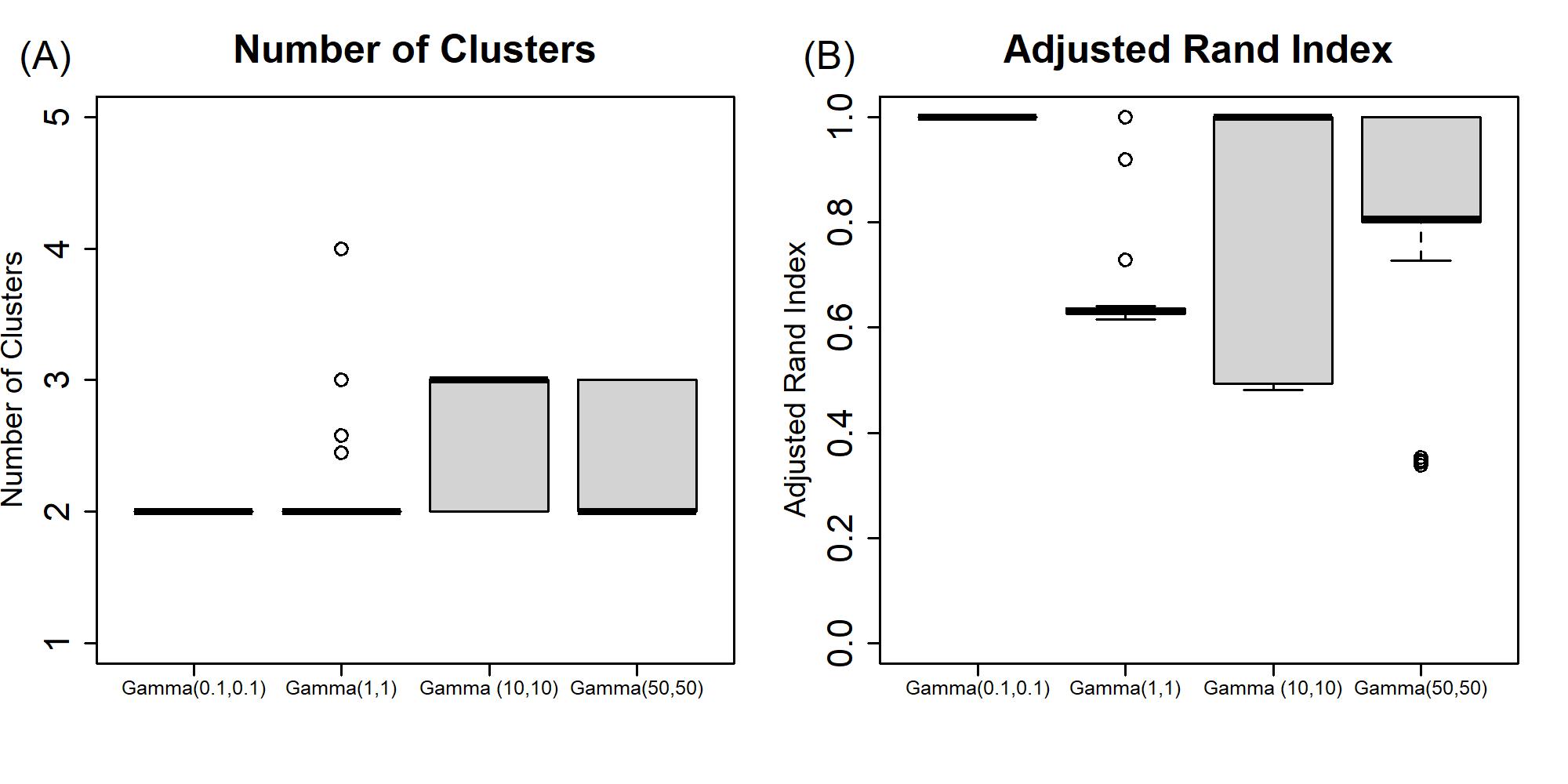}
\caption{Stability of the number of clusters and individual cluster membership over 50 data subsets under different prior for $\alpha$ for the PBC data (Example 2).
(A)  Distribution of the estimated number of clusters by prior distributions of $\alpha$. 
(B) Distribution of the adjusted Rand index by prior distributions of $\alpha$. }
\end{figure}

\newpage
\begin{table}[H]
  \centering
  \caption{List of immune proteins variables included in the cytokine data (Example 1)}
    \scalebox{0.70}{\begin{tabular}{lrl}
    \toprule
    \textbf{Symbol} & \multicolumn{1}{l}{\textbf{Synonym}} & \textbf{Full name} \\
    \midrule
    BDNF  &       & Brain-derived neurotrophic factor \\
    CD40L &       & CD40 ligand \\
    EGF   &       & Epidermal growth factor \\
    ENA78 & \multicolumn{1}{l}{CXCL5 } & Epithelial-derived neutrophil-activating protein 78 \\
    EOTAXIN & \multicolumn{1}{l}{CCL11} &  \\
    FASL  &       & Fas ligand \\
    FGFB  & \multicolumn{1}{l}{FGF2 } & Basic fibroblast growth factor \\
    GCSF  &       & Granulocyte colony-stimulating factor \\
    GMCSF & \multicolumn{1}{l}{CSF2} & Granulocyte-macrophage colony-stimulating factor \\
    GROA  & \multicolumn{1}{l}{CXCL1} & Growth-regulated alpha protein \\
    HGF   &       & Hepatocyte growth factor \\
    ICAM1 &       & Intercellular adhesion molecule 1 \\
    IFNA  &       & Interferon alpha \\
    IFNB  &       & Interferon beta \\
    IFNG  &       & Interferon gamma \\
    IL10  &       & Interleukin-10 \\
    IL12P40 &       & Interleukin-12 P40 \\
    IL12P70 &       & Interleukin-12 P70 \\
    IL13  &       & Interleukin-13 \\
    IL15  &       & Interleukin-15 \\
    IL17A &       & Interleukin-17A \\
    IL17F &       & Interleukin-17F \\
    IL18  &       & Interleukin-18 \\
    IL1A  &       & Interleukin-1 alpha \\
    IL1B  &       & Interleukin-1 beta \\
    IL1RA &       & Interleukin-1 receptor antagonist protein \\
    IL2   &       & Interleukin-2 \\
    IL21  &       & Interleukin-21 \\
    IL22  &       & Interleukin-22 \\
    IL23  &       & Interleukin-23 \\
    IL27  &       & Interleukin-27 \\
    IL31  &       & Interleukin-31 \\
    IL4   &       & Interleukin-4 \\
    IL5   &       & Interleukin-5 \\
    IL6   &       & Interleukin-6 \\
    IL7   &       & Interleukin-7 \\
    IL8   & \multicolumn{1}{l}{CXCL8} & Interleukin-8 \\
    IL9   &       & Interleukin-9 \\
    IP10  & \multicolumn{1}{l}{CXCL10} & Interferon gamma-induced protein 10 \\
    LEPTIN &       & LEPTIN \\
    LIF   &       & Leukemia inhibitory factor \\
    MCP1  & \multicolumn{1}{l}{CCL2} & Monocyte chemoattractant protein 1 \\
    MCP3  & \multicolumn{1}{l}{CCL7} & Monocyte chemoattractant protein 3 \\
    MCSF  & \multicolumn{1}{l}{CSF1} & Macrophage colony-stimulating factor 1 \\
    MIG   & \multicolumn{1}{l}{CXCL9} & Monokine induced by gamma interferon \\
    MIP1A & \multicolumn{1}{l}{CCL3} & Macrophage inflammatory protein-1 alpha  \\
    MIP1B & \multicolumn{1}{l}{CCL4} & Macrophage inflammatory protein-1 beta \\
    NGF   &       & Nerve growth factor \\
    PAI1  & \multicolumn{1}{l}{SERPINE1} & Plasminogen activator inhibitor 1 \\
    PDGFBB & \multicolumn{1}{l}{CSRP2 } & Platelet-derived growth factor-BB \\
    RANTES & \multicolumn{1}{l}{CCL5} & Regulated on Activation, Normal T Cell Expressed and Secreted \\
    RESISTIN & \multicolumn{1}{l}{ADSF} & RESISTIN \\
    SCF   &       & Stem cell factor \\
    SDF1A &       & Stromal cell-derived factor-1 alpha  \\
    TGFA  &       & Transforming growth factor alpha \\
    TGFB  &       & Transforming growth factor beta \\
    TNFA  &       & Tumor necrosis factor alpha \\
    TNFB  &       & Tumor necrosis factor beta \\
    TRAIL &       & TNF-related apoptosis-inducing ligand \\
    VCAM1 &       & Vascular cell adhesion protein 1 \\
    VEGF  &       & Vascular endothelial growth factor A \\
    VEGFD &       & Vascular endothelial growth factor D \\
    CHEX1 &       & Assay CheX beads (Radix BioSolutions) 1 \\
    CHEX2 &       & Assay CheX beads (Radix BioSolutions) 2 \\
    CHEX3 &       & Assay CheX beads (Radix BioSolutions) 3 \\
    CHEX4 &       & Assay CheX beads (Radix BioSolutions) 4 \\
    \bottomrule
    \end{tabular}}
\end{table}%

 \newpage
\section{Supplementary Figures and Tables for the Simulation Study}
\begin{figure}[htbp]  
 \centering
   \includegraphics[width=13cm,height=11cm]{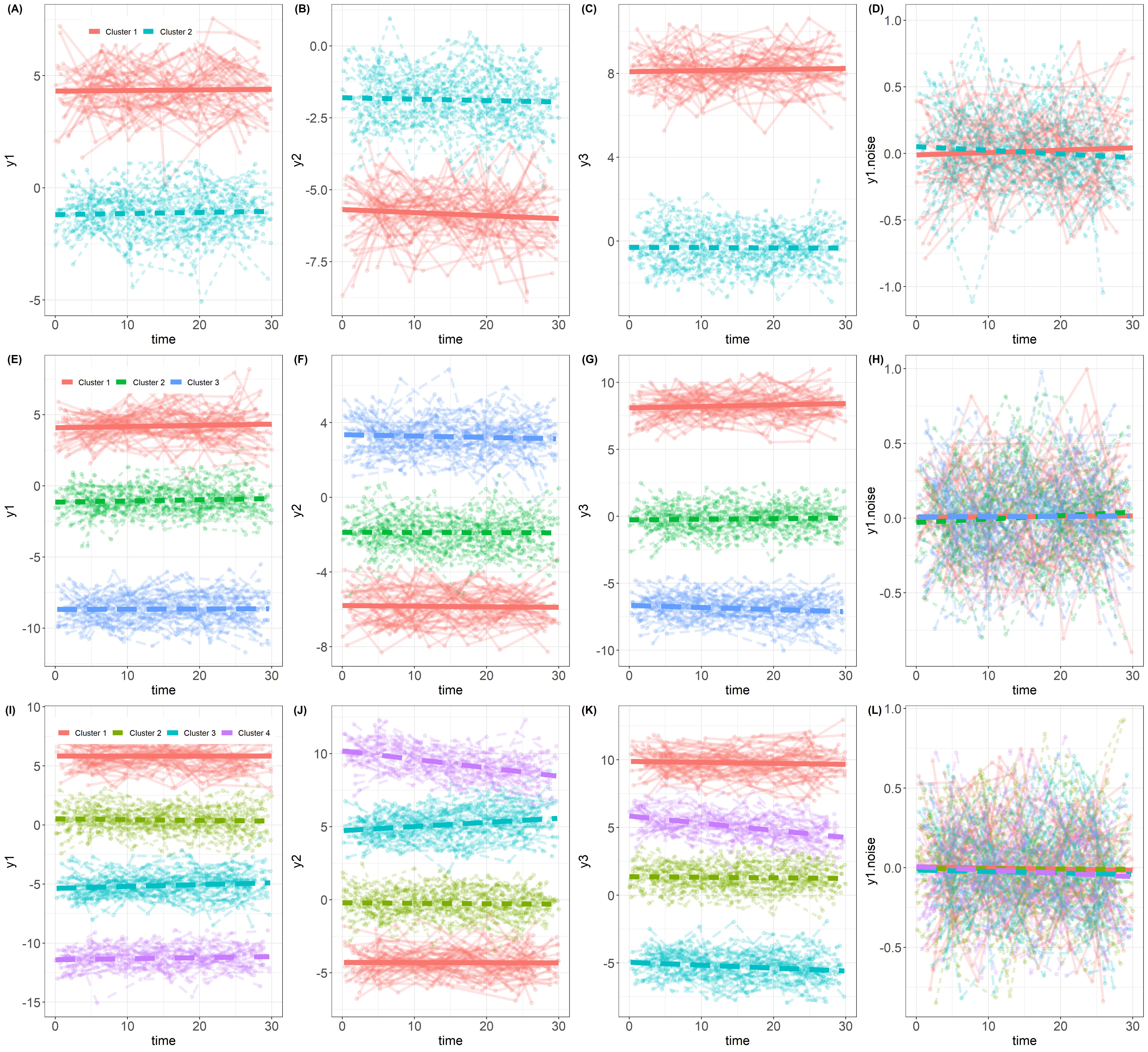}
\caption{Trajectory patterns based on one randomly selected simulated dataset in Scenario 1. (A) Trajectory pattern for $y_1$ under a two-cluster model. (B) Trajectory pattern for $y_2$ under a two-cluster model. (C) Trajectory pattern for $y_3$ under a two-cluster model. (D) Trajectory pattern for $y_{1.\text{noise}}$ under a two-cluster model. (E) Trajectory pattern for $y_1$ under a three-cluster model. (F) Trajectory pattern for $y_2$ under a three-cluster model. (G) Trajectory pattern for $y_3$ under a three-cluster model. (H) Trajectory pattern for  $y_{1.\text{noise}}$ under a three-cluster model.  (I) Trajectory pattern for $y_1$ under a four-cluster model. (J) Trajectory pattern for $y_2$ under a  four-cluster model. (K) Trajectory pattern for $y_3$ under a  four-cluster model. (L) Trajectory pattern for  $y_{1.\text{noise}}$ under a four-cluster model.}
\end{figure}

\newpage
\begin{figure}[htbp]  
 \centering
   \includegraphics[width=14cm,height=12cm]{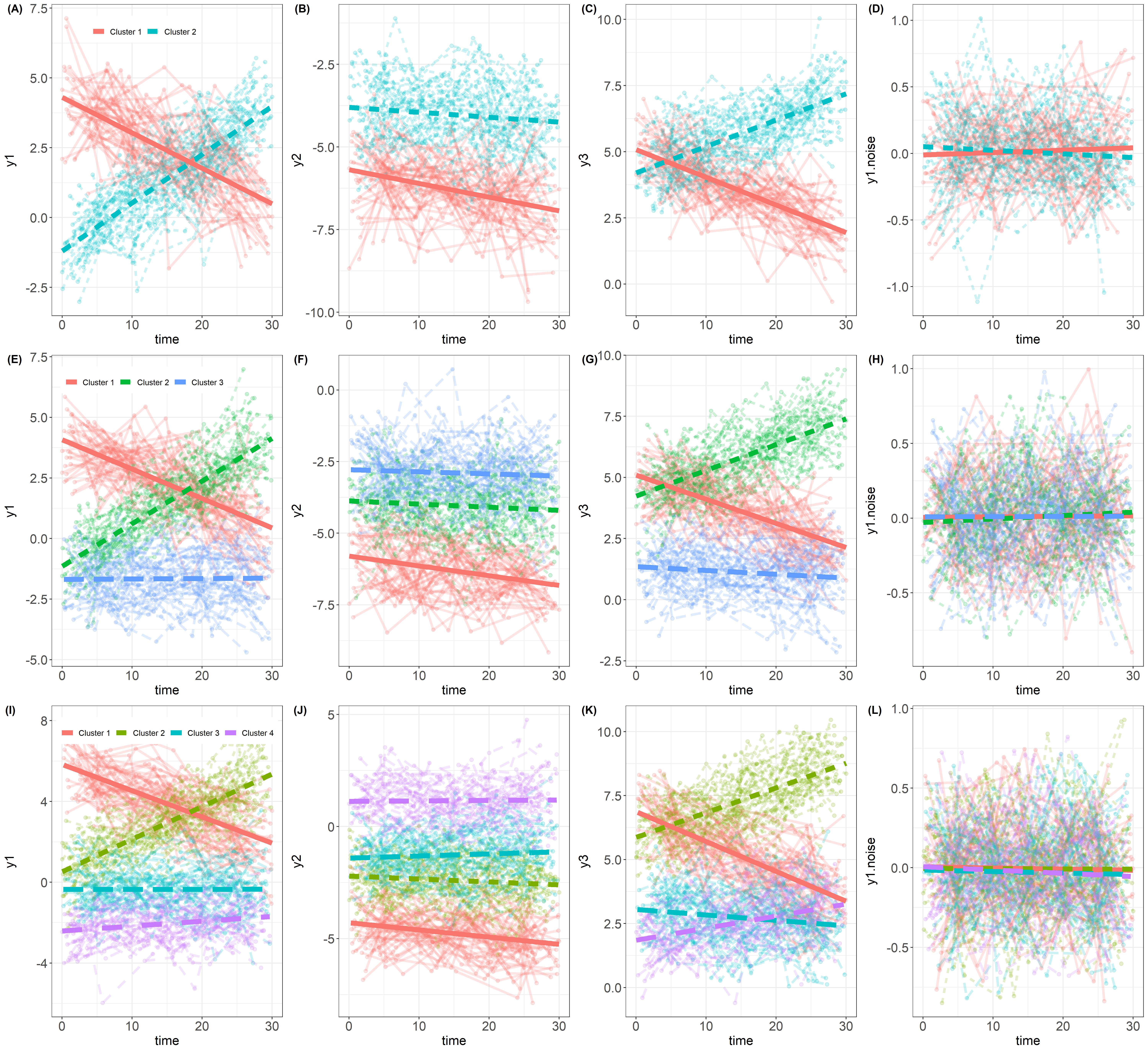}
\caption{Trajectory patterns based on one randomly selected simulated dataset in Scenario 2. (A) Trajectory pattern for $y_1$ under a two-cluster model. (B) Trajectory pattern for $y_2$ under a two-cluster model. (C) Trajectory pattern for $y_3$ under a two-cluster model. (D) Trajectory pattern for $y_{1.\text{noise}}$ under a two-cluster model. (E) Trajectory pattern for $y_1$ under a three-cluster model. (F) Trajectory pattern for $y_2$ under a three-cluster model. (G) Trajectory pattern for $y_3$ under a three-cluster model. (H) Trajectory pattern for  $y_{1.\text{noise}}$ under a three-cluster model.  (I) Trajectory pattern for $y_1$ under a four-cluster model. (J) Trajectory pattern for $y_2$ under a  four-cluster model. (K) Trajectory pattern for $y_3$ under a  four-cluster model. (L) Trajectory pattern for  $y_{1.\text{noise}}$ under a four-cluster model.}
\end{figure}

\newpage
\begin{landscape}
\begin{table}[htbp]
  \centering
  \caption{Mean (SD) of number of clusters and adjusted rand index over 50 simulated datasets for Scenario 1 of the simulation study}
    \scalebox{0.65}{\begin{tabular}{lcccccccccccc}
           \toprule
          & \multicolumn{4}{c}{K = 2}     & \multicolumn{4}{c}{K = 3}     & \multicolumn{4}{c}{K = 4} \\
\cmidrule{2-13}          & 25    & 50    & 75    & 100   & 25    & 50    & 75    & 100   & 25    & 50    & 75    & 100 \\
\cmidrule{2-13}          & \multicolumn{12}{c}{Number of Clusters} \\
    \midrule
    lamb\_long  ($\alpha \sim Gamma(0.1, 0.1)$) & 2(0)  & 2.02(0.14) & 2.12(0.33) & 2.16(0.42) & 3.14(0.57) & 3.06(0.51) & 3.3(0.81) & 3.26(0.56) & 3.98(0.14) & 4.06(0.31) & 3.9(0.51) & 3.48(0.68) \\
    lamb\_long  ($\alpha \sim Gamma(1, 1)$) & 2(0)  & 2.06(0.24) & 2.12(0.39) & 2.12(0.33) & 3.16(0.42) & 3.04(0.4) & 3.38(0.97) & 3.1(0.3) & 4(0.2) & 4.06(0.42) & 3.84(0.51) & 3.46(0.81) \\
    lamb\_long  ($\alpha \sim Gamma(10, 10)$) & 2(0)  & 2.06(0.24) & 2.26(0.56) & 2.16(0.51) & 3.16(0.42) & 3.26(0.85) & 3.4(0.7) & 3.18(0.52) & 3.98(0.14) & 4.02(0.32) & 3.8(0.57) & 3.36(0.66) \\
    lamb\_long  ($\alpha \sim Gamma(50, 50)$) & 2(0)  & 2.04(0.2) & 2.2(0.49) & 2.22(0.42) & 3.22(0.55) & 3(0.29) & 3.36(0.66) & 3.22(0.46) & 4(0.2) & 4.04(0.28) & 3.68(0.47) & 3.34(0.69) \\
    lamb\_twostage ($\alpha \sim Gamma(0.1, 0.1)$) & 1.56(0.5) & 1.42(0.5) & 1.58(0.5) & 1.48(0.5) & 2.9(1.18) & 3.02(1.33) & 2.44(1.07) & 2.02(0.98) & 2.92(1.01) & 2.58(0.91) & 2.58(0.95) & 2.42(0.76) \\
    lamb\_twostage ($\alpha \sim Gamma(1, 1)$) & 1.64(0.48) & 1.52(0.5) & 1.66(0.48) & 1.56(0.5) & 3.24(1.19) & 2.68(1.3) & 2.56(1.7) & 2.86(1.37) & 3.08(1.01) & 2.46(0.84) & 2.36(0.75) & 2.46(0.84) \\
    lamb\_twostage  ($\alpha \sim Gamma(10, 10)$) & 1.68(0.47) & 1.64(0.48) & 1.58(0.5) & 1.62(0.49) & 3.28(1.43) & 3.3(1.43) & 2.9(1.61) & 2.72(1.54) & 3(1.01) & 2.66(0.94) & 2.4(0.81) & 2.44(0.81) \\
    lamb\_twostage  ($\alpha \sim Gamma(50, 50)$) & 1.6(0.49) & 1.52(0.5) & 1.58(0.5) & 1.5(0.51) & 3.44(1.15) & 3.34(1.61) & 2.82(1.41) & 2.86(1.39) & 2.94(1) & 2.6(0.93) & 2.42(0.81) & 2.46(0.84) \\
    lamb\_first($\alpha \sim Gamma(0.1, 0.1)$) & 2(0)  & 2(0)  & 2(0)  & 2(0)  & 3(0)  & 3(0)  & 3.02(0.14) & 3(0)  & 4(0)  & 4.04(0.2) & 4(0)  & 4(0) \\
    lamb\_first ($\alpha \sim Gamma(1, 1)$) & 2(0)  & 2(0)  & 2(0)  & 1.98(0.14) & 3(0)  & 3(0)  & 3(0)  & 3(0)  & 4.02(0.14) & 4.02(0.14) & 4(0)  & 4(0) \\
    lamb\_first  ($\alpha \sim Gamma(10, 10)$) & 2(0)  & 2(0)  & 2(0)  & 2(0)  & 3.04(0.2) & 3(0)  & 3(0)  & 3(0)  & 4.04(0.2) & 4(0)  & 4(0)  & 4.02(0.14) \\
    lamb\_first  ($\alpha \sim Gamma(50, 50)$) & 2(0)  & 2(0)  & 2(0)  & 2(0)  & 3(0)  & 3(0)  & 3(0)  & 3(0)  & 4.02(0.14) & 4(0)  & 4(0)  & 4(0) \\
    lamb\_last ($\alpha \sim Gamma(0.1, 0.1)$) & 2(0)  & 2(0)  & 2(0)  & 1.94(0.24) & 3.02(0.14) & 3(0)  & 3(0)  & 3(0)  & 4(0)  & 4(0)  & 4(0)  & 4(0) \\
    lamb\_last ($\alpha \sim Gamma(1, 1)$) & 2(0)  & 2(0)  & 2(0)  & 1.9(0.3) & 3.04(0.2) & 3(0)  & 3(0)  & 3(0)  & 4(0)  & 4.02(0.14) & 4.02(0.14) & 4(0) \\
    lamb\_last ($\alpha \sim Gamma(10, 10)$) & 2.02(0.14) & 2(0)  & 2(0)  & 1.9(0.3) & 3.02(0.14) & 3(0)  & 3.02(0.14) & 3(0)  & 4(0)  & 4.04(0.2) & 4.02(0.14) & 4(0) \\
    lamb\_last ($\alpha \sim Gamma(50, 50)$) & 2(0)  & 2(0)  & 2(0)  & 1.94(0.24) & 3.04(0.2) & 3.04(0.2) & 3(0)  & 3(0)  & 4(0)  & 4.02(0.14) & 4(0)  & 4(0) \\
    K-means clustering & 5(0)  & 5(0)  & 5(0)  & 2(0)  & 5(0)  & 5(0)  & 5(0)  & 2(0)  & 5(0)  & 5(0)  & 5(0)  & 2(0) \\
    Hierarchical clustering & 2(0)  & 2.02(0.14) & 2.08(0.27) & 2.18(0.44) & 2.18(0.44) & 2.16(0.42) & 2.18(0.44) & 2.24(0.66) & 2.44(0.79) & 2.58(0.88) & 2.46(0.76) & 2.28(0.67) \\
    \midrule
          & \multicolumn{4}{c}{K = 2}     & \multicolumn{4}{c}{K = 3}     & \multicolumn{4}{c}{K = 4} \\
\cmidrule{2-13}          & 25    & 50    & 75    & 100   & 25    & 50    & 75    & 100   & 25    & 50    & 75    & 100 \\
\cmidrule{2-13}          & \multicolumn{12}{c}{Adjusted Rand Index} \\
    \midrule
    lamb\_long  ($\alpha \sim Gamma(0.1, 0.1)$) & 1(0)  & 1(0.03) & 0.99(0.03) & 0.99(0.03) & 0.97(0.09) & 0.98(0.09) & 0.88(0.13) & 0.97(0.06) & 0.99(0.04) & 0.99(0.06) & 0.94(0.11) & 0.81(0.14) \\
    lamb\_long  ($\alpha \sim Gamma(1, 1)$) & 1(0)  & 0.99(0.04) & 0.99(0.03) & 0.99(0.03) & 0.99(0.02) & 0.97(0.1) & 0.9(0.13) & 0.99(0.03) & 0.99(0.04) & 0.98(0.07) & 0.93(0.13) & 0.8(0.13) \\
    lamb\_long  ($\alpha \sim Gamma(10, 10)$) & 1(0)  & 0.99(0.05) & 0.98(0.04) & 0.99(0.03) & 0.99(0.03) & 0.96(0.1) & 0.92(0.09) & 0.98(0.05) & 0.99(0.04) & 0.98(0.06) & 0.92(0.13) & 0.79(0.13) \\
    lamb\_long  ($\alpha \sim Gamma(50, 50)$) & 1(0)  & 0.99(0.04) & 0.98(0.04) & 0.98(0.03) & 0.99(0.04) & 0.97(0.09) & 0.91(0.12) & 0.97(0.06) & 0.99(0.04) & 0.99(0.05) & 0.91(0.14) & 0.78(0.12) \\
    lamb\_twostage ($\alpha \sim Gamma(0.1, 0.1)$) & 0.56(0.5) & 0.42(0.5) & 0.58(0.5) & 0.48(0.5) & 0.78(0.39) & 0.77(0.36) & 0.67(0.44) & 0.51(0.47) & 0.64(0.34) & 0.53(0.3) & 0.53(0.3) & 0.49(0.26) \\
    lamb\_twostage ($\alpha \sim Gamma(1, 1)$) & 0.64(0.48) & 0.52(0.5) & 0.66(0.48) & 0.56(0.5) & 0.86(0.33) & 0.68(0.43) & 0.56(0.45) & 0.8(0.36) & 0.7(0.33) & 0.49(0.28) & 0.46(0.25) & 0.49(0.28) \\
    lamb\_twostage  ($\alpha \sim Gamma(10, 10)$) & 0.68(0.47) & 0.64(0.48) & 0.58(0.5) & 0.62(0.49) & 0.84(0.35) & 0.83(0.34) & 0.69(0.43) & 0.64(0.43) & 0.67(0.33) & 0.56(0.31) & 0.47(0.27) & 0.48(0.27) \\
    lamb\_twostage  ($\alpha \sim Gamma(50, 50)$) & 0.6(0.49) & 0.52(0.5) & 0.58(0.5) & 0.5(0.51) & 0.86(0.31) & 0.79(0.37) & 0.7(0.42) & 0.75(0.39) & 0.65(0.33) & 0.54(0.31) & 0.47(0.27) & 0.5(0.28) \\
    lamb\_first($\alpha \sim Gamma(0.1, 0.1)$) & 1(0)  & 1(0)  & 1(0)  & 1(0)  & 1(0)  & 1(0)  & 1(0)  & 1(0)  & 1(0)  & 1(0)  & 1(0)  & 1(0) \\
    lamb\_first ($\alpha \sim Gamma(1, 1)$) & 1(0)  & 1(0)  & 1(0)  & 0.98(0.14) & 1(0)  & 1(0)  & 1(0)  & 1(0)  & 1(0)  & 1(0)  & 1(0)  & 1(0) \\
    lamb\_first  ($\alpha \sim Gamma(10, 10)$) & 1(0)  & 1(0)  & 1(0)  & 1(0)  & 1(0)  & 1(0)  & 1(0)  & 1(0)  & 1(0)  & 1(0)  & 1(0)  & 1(0) \\
    lamb\_first  ($\alpha \sim Gamma(50, 50)$) & 1(0)  & 1(0)  & 1(0)  & 1(0)  & 1(0)  & 1(0)  & 1(0)  & 1(0)  & 1(0)  & 1(0)  & 1(0)  & 1(0) \\
    lamb\_last ($\alpha \sim Gamma(0.1, 0.1)$) & 1(0)  & 1(0)  & 1(0)  & 0.94(0.24) & 1(0)  & 1(0)  & 1(0)  & 1(0)  & 1(0)  & 1(0)  & 1(0)  & 1(0) \\
    lamb\_last ($\alpha \sim Gamma(1, 1)$) & 1(0)  & 1(0)  & 1(0)  & 0.9(0.3) & 1(0)  & 1(0)  & 1(0)  & 1(0)  & 1(0)  & 1(0)  & 1(0)  & 1(0) \\
    lamb\_last ($\alpha \sim Gamma(10, 10)$) & 1(0)  & 1(0)  & 1(0)  & 0.9(0.3) & 1(0)  & 1(0)  & 1(0)  & 1(0)  & 1(0)  & 1(0)  & 1(0)  & 1(0) \\
    lamb\_last ($\alpha \sim Gamma(50, 50)$) & 1(0)  & 1(0)  & 1(0)  & 0.94(0.24) & 1(0)  & 1(0)  & 1(0)  & 1(0)  & 1(0)  & 1(0)  & 1(0)  & 1(0) \\
    K-means clustering & 1(0)  & 1(0)  & 1(0)  & 1(0)  & 1(0)  & 0.93(0.09) & 0.59(0.08) & 0.5(0.05) & 1(0)  & 0.73(0.07) & 0.59(0.05) & 0.48(0.06) \\
    Hierarchical clustering & 0.99(0.01) & 0.96(0.05) & 0.88(0.08) & 0.78(0.1) & 0.57(0.11) & 0.53(0.11) & 0.5(0.11) & 0.48(0.09) & 0.47(0.13) & 0.47(0.13) & 0.4(0.13) & 0.34(0.11) \\
    \bottomrule
    \end{tabular}}
\end{table}%
 \end{landscape}

\newpage
\begin{landscape}
\begin{table}[htbp]
  \centering
  \caption{Mean (SD) of number of clusters and adjusted rand index over 50 simulated datasets for Scenario 2 of the simulation study}
        \scalebox{0.65}{\begin{tabular}{lcccccccccccc}
       \toprule
          & \multicolumn{4}{c}{K = 2}     & \multicolumn{4}{c}{K = 3}     & \multicolumn{4}{c}{K = 4} \\
\cmidrule{2-13}          & 25    & 50    & 75    & 100   & 25    & 50    & 75    & 100   & 25    & 50    & 75    & 100 \\
    \midrule
    lamb\_long  ($\alpha \sim Gamma(0.1, 0.1)$) & 2(0)  & 2.06(0.24) & 2.04(0.35) & 1.74(0.49) & 3.02(0.14) & 3.14(0.35) & 3.06(0.24) & 3.02(0.38) & 3.72(0.67) & 3.14(0.35) & 3.7(0.81) & 3.62(0.83) \\
    lamb\_long  ($\alpha \sim Gamma(1, 1)$) & 2(0)  & 2.02(0.14) & 2.06(0.24) & 1.64(0.53) & 3(0)  & 3.1(0.3) & 3.08(0.27) & 2.92(0.4) & 3.72(0.67) & 3.1(0.3) & 3.58(0.84) & 3.62(0.92) \\
    lamb\_long  ($\alpha \sim Gamma(10, 10)$) & 2(0)  & 2.04(0.2) & 2.14(0.5) & 1.72(0.67) & 3.02(0.14) & 3.04(0.2) & 3.1(0.3) & 3.04(0.57) & 3.68(0.62) & 3.2(0.4) & 3.56(0.84) & 3.74(0.88) \\
    lamb\_long  ($\alpha \sim Gamma(50, 50)$) & 2(0)  & 2.1(0.3) & 2.02(0.43) & 1.68(0.51) & 3.04(0.2) & 3.04(0.2) & 3.08(0.27) & 2.96(0.35) & 3.76(0.69) & 3.1(0.3) & 3.52(0.79) & 3.82(0.92) \\
    lamb\_twostage ($\alpha \sim Gamma(0.1, 0.1)$) & 1.04(0.2) & 1(0)  & 1(0)  & 1(0)  & 1.56(0.93) & 1.34(0.8) & 1.14(0.5) & 1.06(0.31) & 1.58(0.97) & 1.34(0.92) & 1.28(0.83) & 1.18(0.75) \\
    lamb\_twostage ($\alpha \sim Gamma(1, 1)$) & 1.06(0.31) & 1(0)  & 1(0)  & 1(0)  & 1.74(0.9) & 1.36(0.69) & 1.18(0.48) & 1(0)  & 1.6(0.67) & 1.36(0.63) & 1.32(0.74) & 1.12(0.39) \\
    lamb\_twostage  ($\alpha \sim Gamma(10, 10)$) & 1.08(0.27) & 1.02(0.14) & 1(0)  & 1(0)  & 2.18(1.27) & 1.64(0.88) & 1.06(0.24) & 1.1(0.36) & 1.94(1.08) & 1.54(0.99) & 1.12(0.33) & 1.06(0.24) \\
    lamb\_twostage  ($\alpha \sim Gamma(50, 50)$) & 1.12(0.33) & 1.08(0.44) & 1.04(0.28) & 1(0)  & 2.16(1.08) & 1.42(0.78) & 1.22(0.76) & 1.06(0.24) & 1.68(0.87) & 1.66(1.04) & 1.38(0.88) & 1.16(0.51) \\
    lamb\_first($\alpha \sim Gamma(0.1, 0.1)$) & 1(0)  & 1(0)  & 1(0)  & 1(0)  & 1.16(0.37) & 1.02(0.14) & 1(0)  & 1(0)  & 1.16(0.51) & 1(0)  & 1(0)  & 1(0) \\
    lamb\_first ($\alpha \sim Gamma(1, 1)$) & 1(0)  & 1(0)  & 1(0)  & 1(0)  & 1.14(0.35) & 1(0)  & 1(0)  & 1(0)  & 1.12(0.33) & 1(0)  & 1(0)  & 1.02(0.14) \\
    lamb\_first  ($\alpha \sim Gamma(10, 10)$) & 1(0)  & 1(0)  & 1(0)  & 1(0)  & 1.32(0.47) & 1.02(0.14) & 1(0)  & 1(0)  & 1.1(0.36) & 1(0)  & 1.02(0.14) & 1(0) \\
    lamb\_first  ($\alpha \sim Gamma(50, 50)$) & 1.02(0.14) & 1(0)  & 1(0)  & 1(0)  & 1.24(0.43) & 1(0)  & 1(0)  & 1(0)  & 1.06(0.24) & 1.02(0.14) & 1(0)  & 1(0) \\
    lamb\_last ($\alpha \sim Gamma(0.1, 0.1)$) & 1.02(0.14) & 1(0)  & 1(0)  & 1(0)  & 1.38(0.49) & 1.04(0.2) & 1(0)  & 1(0)  & 1.96(0.2) & 1.58(0.5) & 1.12(0.33) & 1.02(0.14) \\
    lamb\_last ($\alpha \sim Gamma(1, 1)$) & 1.02(0.14) & 1(0)  & 1(0)  & 1(0)  & 1.52(0.5) & 1.04(0.2) & 1(0)  & 1(0)  & 1.98(0.14) & 1.66(0.48) & 1.14(0.35) & 1.02(0.14) \\
    lamb\_last ($\alpha \sim Gamma(10, 10)$) & 1.02(0.14) & 1(0)  & 1(0)  & 1(0)  & 1.54(0.54) & 1(0)  & 1(0)  & 1(0)  & 2(0.2) & 1.56(0.5) & 1.18(0.39) & 1.04(0.2) \\
    lamb\_last ($\alpha \sim Gamma(50, 50)$) & 1.06(0.24) & 1(0)  & 1(0)  & 1(0)  & 1.6(0.53) & 1.02(0.14) & 1(0)  & 1(0)  & 1.96(0.2) & 1.68(0.47) & 1.24(0.43) & 1.02(0.14) \\
    K-means clustering & 5(0)  & 5(0)  & 5(0)  & 2(0)  & 5(0)  & 5(0)  & 5(0)  & 2(0)  & 5(0)  & 5(0)  & 5(0)  & 2(0) \\
    Hierarchical clustering & 2(0)  & 2.68(1.88) & 3.3(2.24) & 5.84(5.66) & 3.14(1.09) & 2.74(0.85) & 2.88(0.94) & 3.66(4.14) & 2(0)  & 2.04(0.2) & 2.04(0.2) & 2.04(0.2) \\
    \midrule
          & \multicolumn{4}{c}{K = 2}     & \multicolumn{4}{c}{K = 3}     & \multicolumn{4}{c}{K = 4} \\
\cmidrule{2-13}          & 25    & 50    & 75    & 100   & 25    & 50    & 75    & 100   & 25    & 50    & 75    & 100 \\
    \midrule
    lamb\_long  ($\alpha \sim Gamma(0.1, 0.1)$) & 1(0)  & 1(0.02) & 0.95(0.2) & 0.68(0.47) & 1(0.01) & 0.98(0.04) & 1(0.02) & 0.97(0.1) & 0.87(0.14) & 0.72(0.07) & 0.78(0.11) & 0.79(0.11) \\
    lamb\_long  ($\alpha \sim Gamma(1, 1)$) & 1(0)  & 1(0.01) & 0.98(0.1) & 0.61(0.48) & 1(0)  & 0.99(0.03) & 0.99(0.02) & 0.95(0.14) & 0.87(0.14) & 0.72(0.06) & 0.78(0.11) & 0.78(0.1) \\
    lamb\_long  ($\alpha \sim Gamma(10, 10)$) & 1(0)  & 1(0.01) & 0.93(0.21) & 0.64(0.47) & 1(0.02) & 1(0.02) & 0.99(0.02) & 0.96(0.12) & 0.88(0.14) & 0.74(0.09) & 0.77(0.1) & 0.79(0.1) \\
    lamb\_long  ($\alpha \sim Gamma(50, 50)$) & 1(0)  & 0.99(0.04) & 0.93(0.24) & 0.65(0.48) & 1(0.02) & 1(0.02) & 0.99(0.02) & 0.96(0.12) & 0.87(0.14) & 0.72(0.04) & 0.74(0.08) & 0.8(0.11) \\
    lamb\_twostage ($\alpha \sim Gamma(0.1, 0.1)$) & 0.04(0.2) & 0(0)  & 0(0)  & 0(0)  & 0.24(0.37) & 0.14(0.28) & 0.06(0.2) & 0.02(0.11) & 0.19(0.25) & 0.1(0.23) & 0.09(0.22) & 0.05(0.17) \\
    lamb\_twostage ($\alpha \sim Gamma(1, 1)$) & 0.04(0.2) & 0(0)  & 0(0)  & 0(0)  & 0.35(0.36) & 0.14(0.24) & 0.09(0.23) & 0(0)  & 0.22(0.24) & 0.13(0.22) & 0.11(0.24) & 0.04(0.14) \\
    lamb\_twostage  ($\alpha \sim Gamma(10, 10)$) & 0.08(0.27) & 0.02(0.14) & 0(0)  & 0(0)  & 0.44(0.4) & 0.3(0.36) & 0.03(0.14) & 0.05(0.19) & 0.27(0.25) & 0.14(0.24) & 0.04(0.13) & 0.02(0.08) \\
    lamb\_twostage  ($\alpha \sim Gamma(50, 50)$) & 0.12(0.33) & 0.04(0.19) & 0.02(0.14) & 0(0)  & 0.45(0.38) & 0.19(0.33) & 0.08(0.26) & 0.03(0.13) & 0.22(0.23) & 0.18(0.26) & 0.11(0.24) & 0.06(0.19) \\
    lamb\_first($\alpha \sim Gamma(0.1, 0.1)$) & 0(0)  & 0(0)  & 0(0)  & 0(0)  & 0.09(0.21) & 0.01(0.08) & 0(0)  & 0(0)  & 0.04(0.11) & 0(0)  & 0(0)  & 0(0) \\
    lamb\_first ($\alpha \sim Gamma(1, 1)$) & 0(0)  & 0(0)  & 0(0)  & 0(0)  & 0.08(0.2) & 0(0)  & 0(0)  & 0(0)  & 0.04(0.11) & 0(0)  & 0(0)  & 0(0) \\
    lamb\_first  ($\alpha \sim Gamma(10, 10)$) & 0(0)  & 0(0)  & 0(0)  & 0(0)  & 0.17(0.26) & 0.01(0.08) & 0(0)  & 0(0)  & 0.02(0.07) & 0(0)  & 0(0)  & 0(0) \\
    lamb\_first  ($\alpha \sim Gamma(50, 50)$) & 0.02(0.14) & 0(0)  & 0(0)  & 0(0)  & 0.13(0.24) & 0(0)  & 0(0)  & 0(0)  & 0.01(0.06) & 0(0)  & 0(0)  & 0(0) \\
    lamb\_last ($\alpha \sim Gamma(0.1, 0.1)$) & 0.02(0.14) & 0(0)  & 0(0)  & 0(0)  & 0.21(0.27) & 0.02(0.11) & 0(0)  & 0(0)  & 0.35(0.1) & 0.19(0.17) & 0.04(0.11) & 0.01(0.05) \\
    lamb\_last ($\alpha \sim Gamma(1, 1)$) & 0.02(0.14) & 0(0)  & 0(0)  & 0(0)  & 0.29(0.28) & 0.02(0.11) & 0(0)  & 0(0)  & 0.35(0.08) & 0.22(0.16) & 0.05(0.12) & 0.01(0.04) \\
    lamb\_last ($\alpha \sim Gamma(10, 10)$) & 0.02(0.13) & 0(0)  & 0(0)  & 0(0)  & 0.29(0.28) & 0(0)  & 0(0)  & 0(0)  & 0.35(0.09) & 0.19(0.17) & 0.06(0.13) & 0.01(0.05) \\
    lamb\_last ($\alpha \sim Gamma(50, 50)$) & 0.06(0.23) & 0(0)  & 0(0)  & 0(0)  & 0.32(0.28) & 0.01(0.08) & 0(0)  & 0(0)  & 0.33(0.08) & 0.23(0.16) & 0.07(0.14) & 0.01(0.05) \\
    K-means clustering & 1(0)  & 1(0)  & 0.99(0.03) & 0.8(0.14) & 1(0)  & 0.96(0.03) & 0.69(0.11) & 0.49(0.05) & 0.74(0.08) & 0.5(0.07) & 0.4(0.05) & 0.39(0.05) \\
    Hierarchical clustering & 0.75(0.1) & 0.59(0.11) & 0.53(0.12) & 0.4(0.18) & 0.66(0.12) & 0.46(0.13) & 0.41(0.11) & 0.36(0.11) & 0.43(0.03) & 0.36(0.04) & 0.33(0.04) & 0.29(0.04) \\
    \bottomrule
    \end{tabular}}%
\end{table}%
 \end{landscape}